\documentclass[%p
reprint,
 amsmath,amssymb,
 aps,
 longbibliography,
 floatfix
]{revtex4-1}

\usepackage{float}
\usepackage{graphicx}% Include figure files
\usepackage{dcolumn}% Align table columns on decimal point
\usepackage{bm}% bold math
\usepackage{hyperref}
\usepackage{CJKutf8}

\usepackage{textgreek} %Greek letter text

\usepackage{etoolbox}

\hypersetup{colorlinks=true,citecolor=blue,linkcolor=black,urlcolor=blue}

\begin{document}

\preprint{APS/123-QED}

\title{Attack and defence in cellular decision-making: lessons from machine learning}

\author{Thomas J. Rademaker \textsuperscript{ 1}, Emmanuel Bengio\textsuperscript{ 2} and Paul Fran\c cois\textsuperscript{ 1}}

\affiliation{\textsuperscript{1}Department of Physics, McGill University, Montreal, Quebec, Canada\\ \textsuperscript{2}School of Computer Science, McGill University, Montreal, Quebec, Canada} %$\dagger$

\date{\today}

\begin{abstract}
 Machine learning algorithms can be fooled by small well-designed adversarial perturbations. This is reminiscent of cellular decision-making where ligands (called antagonists) prevent correct signalling, like in early immune recognition. We draw a formal analogy between neural networks used in machine learning and models of cellular decision-making (adaptive proofreading). We apply attacks from machine learning to simple decision-making models, and show explicitly the correspondence to antagonism by weakly bound ligands. Such antagonism is absent in more nonlinear models, which inspired us to implement a biomimetic defence in neural networks filtering out adversarial perturbations. We then apply a gradient-descent approach from machine learning to different cellular decision-making models, and we reveal the existence of two regimes characterized by the presence or absence of a critical point for the gradient. This critical point causes the strongest antagonists to lie close to the decision boundary. This is validated in the loss landscapes of robust neural networks and cellular decision-making models, and observed experimentally for immune cells. For both regimes, we explain how associated defence mechanisms shape the geometry of the loss landscape, and why different adversarial attacks are effective in different regimes. Our work connects evolved cellular decision-making to machine learning, and motivates the design of a general theory of adversarial perturbations, both for \textit{in vivo} and \textit{in silico} systems.
\end{abstract}

\pacs{Valid PACS appear here}% PACS, the Physics and Astronomy
                             % Classification Scheme.
%\keywords[showkeys]{Suggested keywords}%Use showkeys class option if keyword
                              %display desired
\maketitle

%\tableofcontents

\section{INTRODUCTION}
\begin{figure*}[!thb]
\centering
\includegraphics[width=0.75\linewidth]{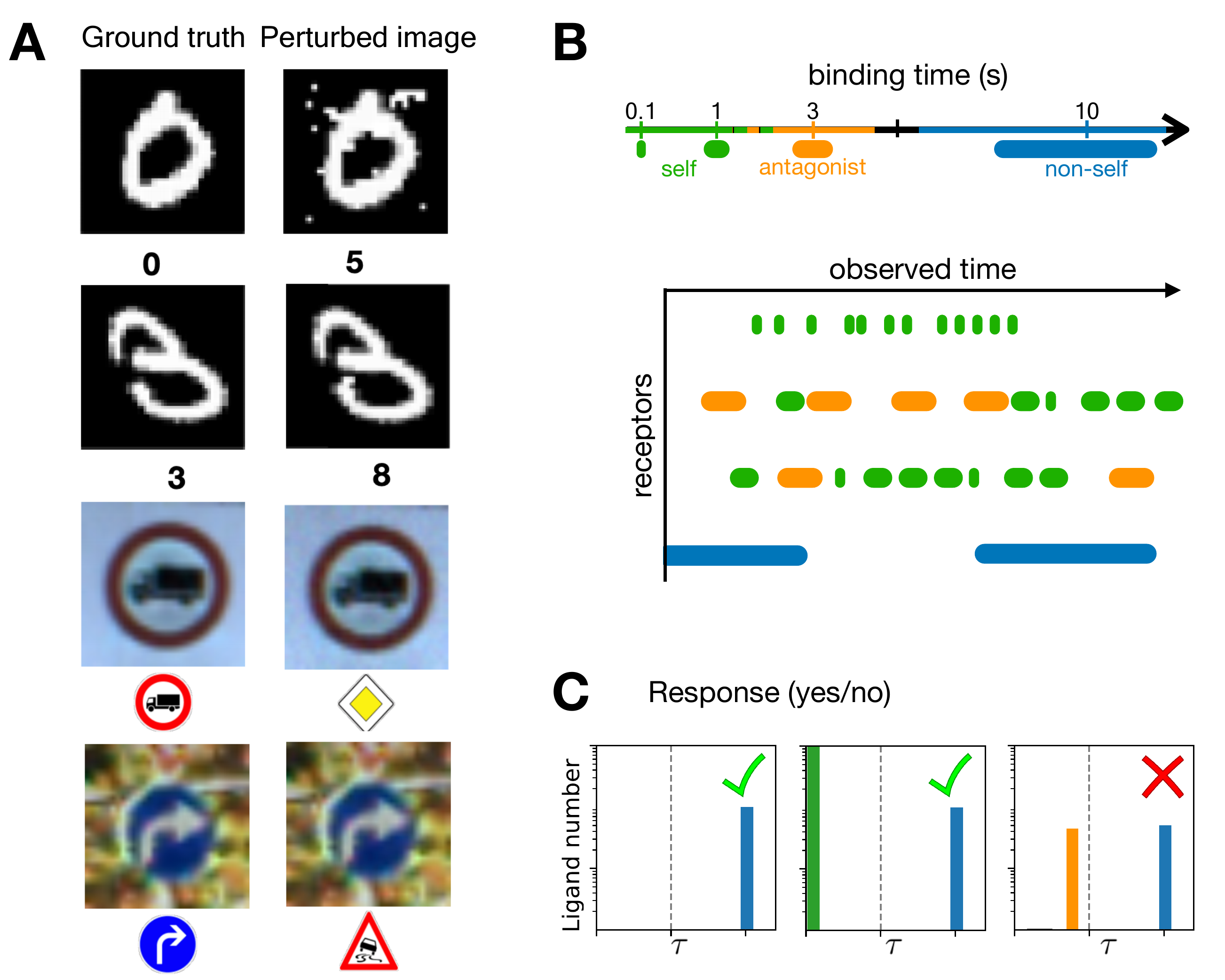}%
\caption{
  \textbf{Ligand discrimination and digit recognition tasks} (A) Adversarial examples on digits and roadsigns. Reproduced from \cite{Papernot2017}. Left column displays original images with categories recognized by machine learning algorithms, right column displays images containing targeted perturbations leading to misclassification. (B) Schematics of ligand binding events showing typical receptor occupancy through some observed time during cellular decision-making using T cell terminology (``self vs non-self"). The colored bars corresponds to self (green), antagonist (orange) and non-self (blue) ligands binding to receptors. Their lengths are indicative of the binding time $\tau_i$, whereas their rate of binding measures the on-rate $k^\text{on}_i$. (C) Different ligand distributions give different response. The vertical dotted line indicates quality $\tau_d$. Decision should be to activate if one observes ligands with $\tau>\tau_d$, so on the right of the dotted line. In an immune context, T cells responds to ligand distributions of agonists alone and agonists in the presence of non-agonists (with very small binding times $\tau$), while the T cell fails to respond if there are too many ligands just below threshold $\tau_d$. }
\label{fig:tasks}
\end{figure*}

\vspace{-10pt}
Machine learning is becoming increasingly popular with major advances coming from deep neural networks \cite{LeCun2015}. Deep learning has improved the state-of-the-art in automated tasks like image processing \cite{Krizhevsky2012}, speech recognition \cite{Hinton2012} and machine translation \cite{Sutskever2014}, and has already seen a wide range of applications in research and industry. Despite their success, neural networks suffer from blind spots: small perturbations added to unambiguous samples may lead to misclassification \cite{Szegedy2013}. Such adversarial examples are most obvious in image recognition, for example, a panda is misclassified as a gibbon or a handwritten 3 as a 7 \cite{Goodfellow2014}. Real world scenarios exist, like adversarial road signs fooling computer vision algorithms (Fig.~\ref{fig:tasks} A) \cite{Papernot2017}, or adversarial perturbations on medical images triggering incorrect diagnosis \cite{Finlayson2018}. Worse, adversarial examples are often transferable across algorithms (see \cite{Akhtar2018} for a recent review), and certain universal perturbations fool any algorithm. \cite{Moosavi-Dezfooli2017}. 

Categorization and inference are also tasks found in cellular decision-making \cite{Siggia2013}.  For instance, T cells have to discriminate between foreign and self ligands which is challenging since foreign ligands might not be very different biochemically from self ligands \cite{Gascoigne2001,Feinerman2008}. Decision-making in an immune context is equally prone to detrimental perturbations in a phenomenon called ligand antagonism \cite{Francois2016a}. Antagonism appears to be a general feature of cellular decision-makers: it has been observed in T cells \cite{Altan-Bonnet2005}, mast cells \cite{Torigoe1998} and other recognition processes like olfactory sensing \cite{Reddy2018, Tsitron2011}.

There is a natural analogy to draw between decision-making in machine learning and in biology. In machine learning terms, cellular decision-making is similar to a classifier. Furthermore, in both artificial and cellular decision-making, targeted perturbations lead to faulty decisions even in the presence of a clear ground truth signal. As a consequence, arms races are observed in both systems. Mutating agents might systematically explore ways to fool the immune cells via antagonism, as has been proposed in the HIV case \cite{Klenerman1994,Meier1995,Kent1997}. Recent examples might include neoantigens in cancer \cite{Snyder2014,Schumacher2015} which are implicated in tumour immunoediting and escape from the immune system. Those medical examples are reminiscent of how adversaries could generate black box attacks aimed at fooling neural networks \cite{Papernot2017}. Strategies for provable defenses or robust detection of adversarial examples \cite{Grosse2017,Wong2018} are currently developed in machine learning, but we are still far from a general solution.

In the following, we draw a formal correspondence between biophysical models of cellular decision-making displaying antagonism on the one hand, and adversarial examples in machine learning on the other hand. We show how simple attacks in machine learning mathematically correspond to antagonism by many weakly bound ligands in cellular decision-making. Inspired by kinetic proofreading in cellular decision-making, we implement a biomimetic defence for digit classifiers, and we demonstrate how these robust classifiers exhibit similar behavior to the nonlinear adaptive proofreading models. Finally, we explore the geometry of the decision boundary for adaptive proofreading, and observe how a  critical point in the gradient dynamics emerges in networks robust to adversarial perturbations. Recent findings in machine learning \cite{Krotov2016} confirm the existence of two regimes, which are separated by a large nonlinearity in the activation function. This inspired us to define two categories of attack (high-dimensional, small amplitude and low-dimensional, large amplitude) both for models of cellular decision-making and neural networks. Our work suggests the existence of a unified theory of adversarial perturbations for both evolved and artificial decision-makers.

\subsection{Adaptive proofreading for cellular decision-making}
\vspace{-10pt}
Cellular decision-making in our context refers to classification of biological ligands in two categories, e.g.  ``self vs non-self" in immunology, or ``agonist vs non-agonist" in physiology \cite{Feinerman2008,Das2010,Lagarde2015}. For most of those cases, qualitative distinctions rely on differences in a continuously varying property (typically a biochemical parameter). Thus it is convenient to rank different ligands based on a parameter (notation $\tau$) that we will call quality. Mathematically, a cell needs to decide if it is exposed to ligands with quality $\tau > \tau_d$, where $\tau_d$ is  the quality at the decision threshold. Such ligands triggering response are called agonists. A general problem then is to consider cellular decision-making based on ligand quality  irrespective of ligand quantity (notation $L$).  An example can be found in immune recognition with the lifetime dogma \cite{Feinerman2008}, where it is assumed that a T cell discriminates ligands based on their characteristic  binding time $\tau$ to T cell receptors (this is of course an approximation and other parameters might also play a role in defining quality, see \cite{ Govern2010,Chakraborty2014,Lever2016}). Ligand discrimination is a nontrivial problem for the cell, which does not measure single-binding events but only has access to global quantities such as the total number of bound receptors (Fig.~\ref{fig:tasks} B).  The challenge is to ignore many subthreshold ligands ($\tau<\tau_d$) while responding to few agonist ligands with $\tau > \tau_d$ \cite{Altan-Bonnet2005,Feinerman2008,Francois2013}. In particular, it is known experimentally in many different contexts that addition of antagonistic subthreshold ligands can impair proper decision-making (Fig.~\ref{fig:tasks} C) \cite{Altan-Bonnet2005,Torigoe1998,Reddy2018}.

\begin{figure*}[!thb]
\centering
\includegraphics[width=0.75\linewidth]{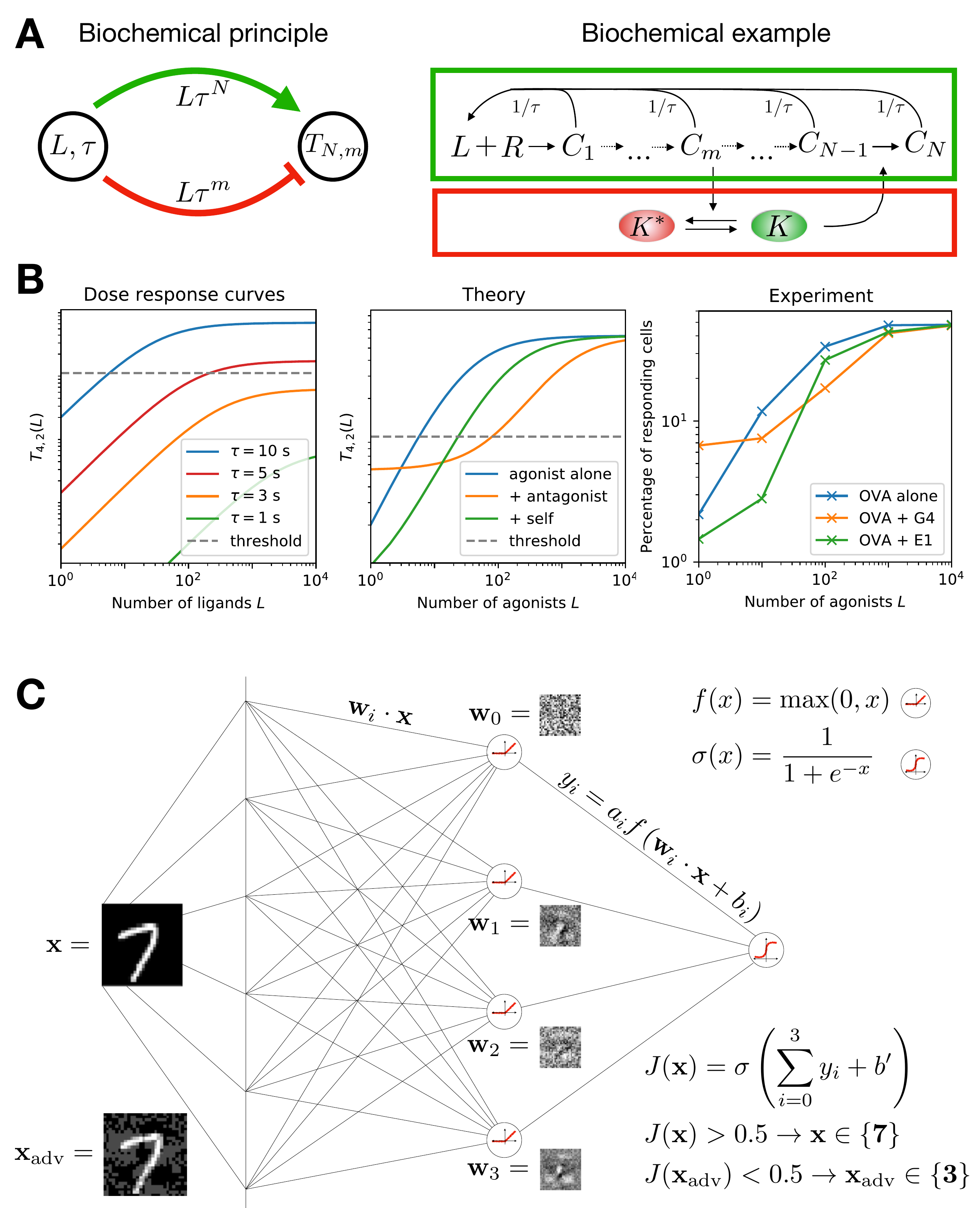}%
\caption{\textbf{Adaptive proofreading and neural network} (A) Left: Adaptive proofreading networks have an activating and repressing branch with different weights on $\tau$. Right: detailed adaptive proofreading network adapted from \cite{Francois2016b}. Ligand $L$ binds to receptor $R$ to form unphosphorylated complex $C_0$. The receptor chain is iteratively phosphorylated until reaching state $C_N$ along the activating branch (green). At every stage $C_i$, the ligand can unbind from the receptor with ligand-specific rate $\tau^{-1}$. At $C_m$, the repressing branch (red) splits by inhibiting the kinase $K$, which mediates the feedforward mechanism. (B) Dose-response curves for pure ligand types and mixtures, in both adaptive proofreading models and experiments on T cells (redrawn from \cite{Francois2013}). Details on models and parameters used are given in Appendix S2. For experiments, OVA are agonist ligands, G4 and E1 are ligands known to be below threshold, but showing clear antagonistic properties. (C) Schematic of the neural network used for digit recognition. We explicitly show the 4 weight vectors $W_i$ learned in one instance of the training, the activation function $J$ and an adversarially perturbed sample $\mathbf{x}_\text{adv}$.)
}

\label{fig:networks}
\end{figure*}

To model cellular decision-making, we will use the general class of ``adaptive sorting'' or ``adaptive proofreading'' models, which account for many aspects of immune recognition \cite{Lalanne2013,Francois2016a}, and can be shown to capture all relevant features of such cellular decision-making close to a decision threshold \cite{Francois2016b}.  An example of such a model is displayed in Fig.~\ref{fig:networks} A. Importantly, we have shown previously that many other biochemical models present similar properties for the steady-state response as a function of the input ligand distribution \cite{Proulx-Giraldeau2017}. In the following we summarize the most important mathematical properties of such models. An analysis of the detailed biochemical kinetics of the model of Fig.~\ref{fig:networks} A is presented in Appendix S1.

%\begin{figure}[t]
%\justify
%  \contcaption{\textbf{Adaptive proofreading and neural network} (A) Left: Adaptive proofreading networks have an activating and repressing branch with different weights on $\tau$. Right: detailed adaptive proofreading network adapted from \cite{Francois2016b}. Ligand $L$ binds to receptor $R$ to form unphosphorylated complex $C_0$. The receptor chain is iteratively phosphorylated until reaching state $C_N$ along the activating branch (green). At every stage $C_i$, the ligand can unbind from the receptor with ligand-specific rate $\tau^{-1}$. At $C_m$, the repressing branch (red) splits by inhibiting the kinase $K$, which mediates the feedforward mechanism. (B) Dose-response curves for pure ligand types and mixtures, in both adaptive proofreading models and experiments on T cells (redrawn from \cite{Francois2013}). Details on models and parameters used are given in Appendix S2. For experiments, OVA are agonist ligands, G4 and E1 are ligands known to be below threshold, but showing clear antagonistic properties. (C) Schematic of the neural network used for digit recognition. We explicitly show the 4 weight vectors $W_i$ learned in one instance of the training, the activation function $J$ and an adversarially perturbed sample $\mathbf{x}_\text{adv}$.}
%\end{figure}

We assume an idealized situation where a given receptor $i$, upon ligand binding (on-rate $k^\text{on}_i$, binding time  $\tau_i$) can exist in $N$ biochemical states (corresponding to phosphorylation stages of the receptor tails in the immune context \cite{McKeithan1995,Kersh1998}). Those states allow the receptor to effectively compute different quantities, such as $c^i_n=k^\text{on}_i \tau_i^n, 0 \leq n \leq N$, which can be done with kinetic proofreading \cite{Hopfield1974,Ninio1975,McKeithan1995}. In particular, ligands with larger $\tau$ give a relatively larger value of $c^i_N$ due to the geometric amplification associated with proofreading steps. We assume receptors to be identical, so that any downstream receptor processing by the cell must be done on the sum(s) $C_n=\sum_i c^i_n=\sum_i k^\text{on}_i \tau_i^n$. We also consider a quenched situation in which only one ligand is locally available for binding to every receptor. In reality, there is a constant motion of ligands, such that $k^\text{on}_i$ and $\tau_i$ are functions of time and stochastic treatments are required \cite{Siggia2013, Lalanne2015, Mora2015}, but on the time-scale of primary decision-making it is reasonable to assume that the ligand distribution does not change much \cite{Altan-Bonnet2005}.

Adaptive proofreading models rely on an incoherent feedforward loop, where an output is at the same time activated and repressed by bound ligands via two different branches in a biochemical network (Fig.~\ref{fig:networks} A). An explicit biochemical example is shown on the right panel of Fig.~\ref{fig:networks} A. Here, activation occurs through a kinetic proofreading cascade (green arrow/box), and repression through the inactivation of a kinase by the same cascade (red arrow/box). The branches engage in a tug-of-war, which we describe below.

For simplicity, let us first assume that only one type of ligands with binding time $\tau$ and on rate $k_\text{on}$ are presented. We call $L$ the quantity of ligands. Then, in absence of saturation, the total number of $n$-th complex $C_n$ of the proofreading cascade along the activation branch will be proportional to $k_\text{on}L \tau^n$. This is the activation part of the network where the response is activated.

We now assume that the $m$-th complex of the cascades are inactivating a kinase $K$ specific to $C_m$, so that $K\propto (k_\text{on}L \tau^m)^{-1}$ for $L$ big enough. This is the repression part of the network. $K$ is assumed to diffuse freely and rapidly between receptors so that it effectively integrates information all over the cell (recent work quantified how this crosstalk can indeed improve detection \cite{Carballo2018}). $m$ is an important parameter that we will vary to compare different models. $K$ then catalyzes the phosphorylation of the final complex of the cascade so that we have for the total number  $C_N$ 
\begin{equation}
\dot{C}_N = K C_{N-1} - \tau^{-1} C_N. \label{C_N}
\end{equation}

and at steady state
\begin{equation}
C_N\propto \frac{k^\text{on} L \tau^N}{k^\text{on} L \tau^m} = \tau^{N-m}
\end{equation}

The $L$ dependence cancels, and $C_N$ is a function of $\tau$ alone. From this, it is clear that ligand classification can be done purely based on $C_N$, the total number of complexes, which is a measure of ligand quality. In this situation, it is easy to define a threshold $\tau_d^{N-m}$ that governs cell activation $(C_N > \tau_d^{N-m})$ or quiescence $(C_N < \tau_d^{N-m})$. Biochemically, this can be done via the digital activation of another kinase shared between all receptors \cite{Altan-Bonnet2005,Lalanne2013}.

This model can be easily generalized to a mixture of ligands with different qualities. To do so, in the previous derivations all quantities accounting for the total  complex $C_n$ of the form $k_\text{on}L\tau^n$ can be replaced by $\sum_i k^\text{on}_i L_i \tau_i^N$, calling $L_i$ the quantity of ligands with identical $k^\text{on}_i, \tau_i$. We then define the generalized output of the biochemical network as
\begin{equation}
T_{N,m} = \frac{\sum_i k^\text{on}_i L_i \tau_i^N}{\sum_i k^\text{on}_i L_i \tau_i^m}. \label{generalTnm}
\end{equation}
Similar equations for an output $T_{N,m}$ can be derived for many types of networks, as described in \cite{Proulx-Giraldeau2017}. For this reason we will focus in the following on the properties of $T_{N,m}$, forgetting about the internal biochemistry giving rise to this behaviour. Notice here that by construction $N>m>1$, but other cases are posssible with different biochemistry, for instance examples in olfaction correspond to the case $N=1,m=0$ \cite{Reddy2018} (see also another example in \cite{Francois2016b})  . Also notice that if kinetic parameters of the ligands are not identical, the dependence on $L_i$ does not cancel out, which will be the origin of most of the key phenomena described below.

%In this situation, the activation and repression branch are assumed to be activated linearly as a function of $L$. We further assume that the $\tau$ dependency of the activation branch is stronger (parameter $N$, assumed to be an integer representing number of phosphorylation sites on the T cell receptor chain in immune recognition models) than the repression branch (parameter $m$, with $m<N$). 
%If $L$ ligands with identical $\tau$ and $k^\text{on}$ are presented to a cell, we have $T_{N,m}=\frac{k^\text{on} L \tau^N}{k^\text{on} L \tau^m} = \tau^{N-m}$. 
%The output variable of the biochemical network $T_{N,m}$ is given by the summation over all ligands $i$ of all N-times phosphorylated complexes $C_N^i $, which is the positive part of the network. The stationary solution for $T_{N,m}$ then is

Fig.~\ref{fig:networks} B shows theoretical and experimental curves of a realistic adaptive proofreading model (including minimum concentration for repression of kinase $K$, etc. see Appendix S2 for full model and parameter values). We have chosen $(N,m) = (4,2)$ so that the qualitative features of the theoretical curves match the experimental curves best. Adaptive proofreading models give dose response curves plateauing at different values as a function of parameter $\tau$, allowing to perform sensitive and specific measurement of this parameter. For small $\tau$ (e.g. $\tau=3 \, $s), one never reaches the detection threshold (dotted line on Fig.~\ref{fig:networks} B, left panel) even for many ligands. For slightly bigger $\tau = 10 \, \textrm{s} > \tau_\text{d}$, the curve is shifted up so that detection is made even for a small concentration of agonists.

Nontrivial effects appear if we consider mixtures of ligands with different qualities. Then the respective computation made by the activation and repression branch of the network depend in different ways on the distribution of the presented ligand binding times. For instance, if we now add $L_\text{a}$ antagonists with lower binding time $\tau_\text{a}<\tau$ and equal on-rate $k^\text{on}$, we have $T_{N,m}=\frac{L\tau^N+L_\text{a}\tau_\text{a}^N}{L\tau^m+L_\text{a}\tau_\text{a}^m}$,  which is smaller than the response $\tau^{N-m}$ for a single type of ligands, corresponding to ligand antagonism (Fig.~\ref{fig:networks} B, middle panel) \cite{Germain1999,Dittel1999,Altan-Bonnet2005,Francois2016a}. In the presence of many ligands below the threshold of detection, the dose response curve  are simultaneously moved to the right but with a higher starting point (compared to the reference curve for ``agonist alone''), as observed experimentally (Fig.~\ref{fig:networks} B, right panel, data redrawn from \cite{Francois2013}). Different models have different antagonistic properties, based on the strength of the activation branch ($N$) relative to the repression branch ($m$). More mathematical details on these models can be found in \cite{Lalanne2013,Francois2016a,Francois2016b}.

%and antagonism ( compared to a model similar to Eq. \ref{Output} with $(N,m)=(4,2)$ and identical on-rates 

%Antagonistic effects thus occur when the repression branch is activated more strongly than the activation branch (with respect to a reference situation), thus killing the response. Fig.~\ref{fig:networks} C middle and right panels illustrate this.

\vspace{-10pt}
\subsection{Neural networks for artificial decision-making}
\vspace{-10pt}
We will compare cellular decision-making to decision-making in machine learning algorithms. We will constrain our analysis to binary decision-making (which is of practical relevance, for instance in medical applications \cite{Finlayson2018}), using as a case-study image classification from two types of digits. These images are taken from MNIST \cite{Lecun1998}, a standard database with 70000 pictures of handwritten digits. Even for such a simple task, designing a good classifier is not trivial, since it should be able to classify irrespective of subtle changes in shapes, intensity and writing style (i.e. with or without a central bar for a $7$).

A simple machine learning algorithm is logistic regression. Here, the inner product of the input and a learned weight vector determines the class of the input. Another class of machine learning algorithms are feedforward neural networks: interconnected groups of nodes processing information layer-wise. We chose to work with neural networks for several reasons. First, logistic regression is a limiting case of a neural network without hidden layers. Second, a neural network with one hidden layer more closely imitates information processing in cellular networks, i.e. in the summation over multiple phosphorylation states of the receptor-ligand complex (nodes) in a biochemical network. Third, such an architecture reproduces classical results on adversarial perturbations such as the ones described in \cite{Goodfellow2014}. Fig.~\ref{fig:networks} C introduces the iterative matrix multiplication inside a neural network. Each neuron $i$ computes $\mathbf{w}_i \cdot \mathbf{x}, \, i \in [0,3]$, adds bias $b_i$, and transforms the result with an activation function $f(x)$. We chose to use a Rectified Linear Unit (ReLU), which returns 0 when its input is negative, and the input itself otherwise. The resulting $f(\mathbf{w}_i \cdot \mathbf{x} + b_i)$ is multiplied by another weight vector with elements $a_i$, summed up with a bias, defining a scalar quantity $x=\sum_i a_i f(\mathbf{w}_i \cdot \mathbf{x} + b_i) + b'$. Finally, we obtain the score $J(\mathbf{x})$ (a probability between 0 and 1 for the input to belong to a class) by transforming $x$ with the logistic function $\sigma(x)$. Parameters of such networks are optimized using classical stochastic gradient descent within a scikit implementation \cite{scikit-learn}, see Appendix S2. As an example, in Fig.~\ref{fig:networks} C, a 7 is correctly classified by the neural network ($J(\mathbf{x}) > 0.5$), while the adversarial 7 is classified as a three ($J(\mathbf{x}_\text{adv}) < 0.5$).

\vspace{-10pt}
\section{RESULTS}
\vspace{-10pt}

We first summarize the general approach followed to draw the parallel between machine learning and cellular decision-making.  We will limit ourselves to simple classifications where a single decision is made, such as ``agonist present vs no agonist present'' in biology, or ``3 vs 7'' in digit recognition. As input samples, we will consider pictures in machine learning, and ligand distributions in biology. We define a ligand distribution as the set of concentrations with which the ligands with unique binding times are present. This corresponds to a picture that is presented as a histogram of pixel values; the spatial correlation between pixels is lost, but their magnitude remains preserved. Decision-making on a sample is then done via a scoring function (or score). This score is computed either directly by the machine learning algorithm (score $J$) or by the biochemical network, via the concentration of a given species (score $T_{N,m}$). For simple classifications, the decision is then based on the relative value of the score above or below some threshold (typically $0.5$ for neural networks where decision is based on sigmoidal functions, or some fixed value related to the decision time $\tau_d$ for biochemical networks). 

The overall performance of a given classifier depends on the behavior of the score in the space of possible samples (i.e. the space of all possible pictures, or the space of all possible ligand distributions). Both spaces have high dimensions: for instance the dimension in the MNIST picture correspond to number of pixels $28\times28= 784$, while in immunology ligands can bind to roughly $30000$ receptors \cite{Altan-Bonnet2005}. The score can thus be thought of as a nonlinear projection of this high-dimensional space in one dimension. We will study how the score behaves in relevant directions in the sample space, and how to change the corresponding geometry and position of decision boundaries (defined as the samples where the score is equal to the classification threshold). We will show that similar properties are observed, both close to typical samples and to the decision boundary. It is important to notice at this stage that the above considerations are completely generic on the biology side and are not necessary limited to, say immune recognition. However, we will show that adaptive proofreading presents many features reminiscent of what is observed in machine learning.

\vspace{-10pt}
\subsection{Fast Gradient Sign Method recovers antagonism by weakly binding ligands}
\vspace{-10pt}

In this framework, from a given sample, an adversarial perturbation is a small perturbation in sample space giving a change in score reaching (or crossing) the decision boundary. We start by mathematically connecting the simplest class of adversarial examples in machine learning to antagonism in adaptive proofreading models. We follow the original Fast Gradient Sign Method (FGSM) proposed by \cite{Goodfellow2014}. The FGSM computes the  local maximum adversarial perturbation $\eta=\epsilon \, \text{sgn} \left(\nabla_x J \right)$ (where sign is taken elementwise). $\nabla_x J$ represents the gradient of the scoring function, categorizing images in two different categories (such as 3 and 7 in \cite{Goodfellow2014}). Its elementwise sign defines an image, that is added to the initial batch of images with small weight $\epsilon$. Examples of such perturbations are shown in Fig.~\ref{fig:networks} C (bottom left) and Fig.~S2 A for the 3 vs 7 digit classification problem. While to the human observer, the perturbation is weak and only changes the background, naive machine learning algorithms are completely fooled by the perturbation and systematically misclassify the digit. 

%with $||\eta||_{\infty}\leq\epsilon$.

Coming back to adaptive proofreading models, we apply FGSM for the computation of a maximally antagonistic perturbation. To do so, we need to specify the equivalent of pixels in adaptive proofreading models. A natural choice is to consider parameters associated to each pair (index $i$) of receptor/ligands, namely $k^\text{on}_i$ (corresponding to the rate at which ligands bind to receptors, also called on-rate \footnote{The on-rate is easily confused with the unbinding rate, whose inverse we call the binding time, which indicates the lifetime of the ligand-receptor complex}) and $\tau_i$ (corresponding to quality). If a receptor $i$ is unoccupied, we set its $k_i$ and $\tau_i$ to $0$ \footnote{an alternative choice without loss of generality is to consider a situation where for unoccupied receptors, $k_i$ is $0$ but $\tau_i$ is arbitrary, corresponding to a ligand available for binding}. We then compute gradients with respect to these parameters.
%This choice makes biological sense since the biological classifier should detect ligands of high quality within a sea  of available ligands with very small binding times/quality. % to be the majority of potential ligands.

As a simple example, we start with the case  $(N,m)=(1,0)$, which also corresponds to a recently proposed model for antagonism in olfaction \cite{Reddy2018}, with the role of $k^\text{on}$ played by inverse affinity $\kappa^{-1}$, the role of $\tau$ played by efficiency $\eta$, and the spiking rate of the olfactory receptor neurons is $J(T_{N,m})$, that can be interpreted as a scoring function in the machine learning sense. In this case, $T_{1,0}$ simply computes the average quality  $\tau_\text{avg}$ of ligands presented weighted by $k^\text{on}_i$ (models with $N>m>0$ give less intuitive results as will be shown in the following). It should be noted that while this computation is formally simple, biochemically it requires elaborated internal interactions, because a cell can not easily disentangle influence of individual receptors, see \cite{Francois2016a,Reddy2018} for explicit examples.

\begin{figure*}[!thb]
\centering
\includegraphics[width=0.75\linewidth]{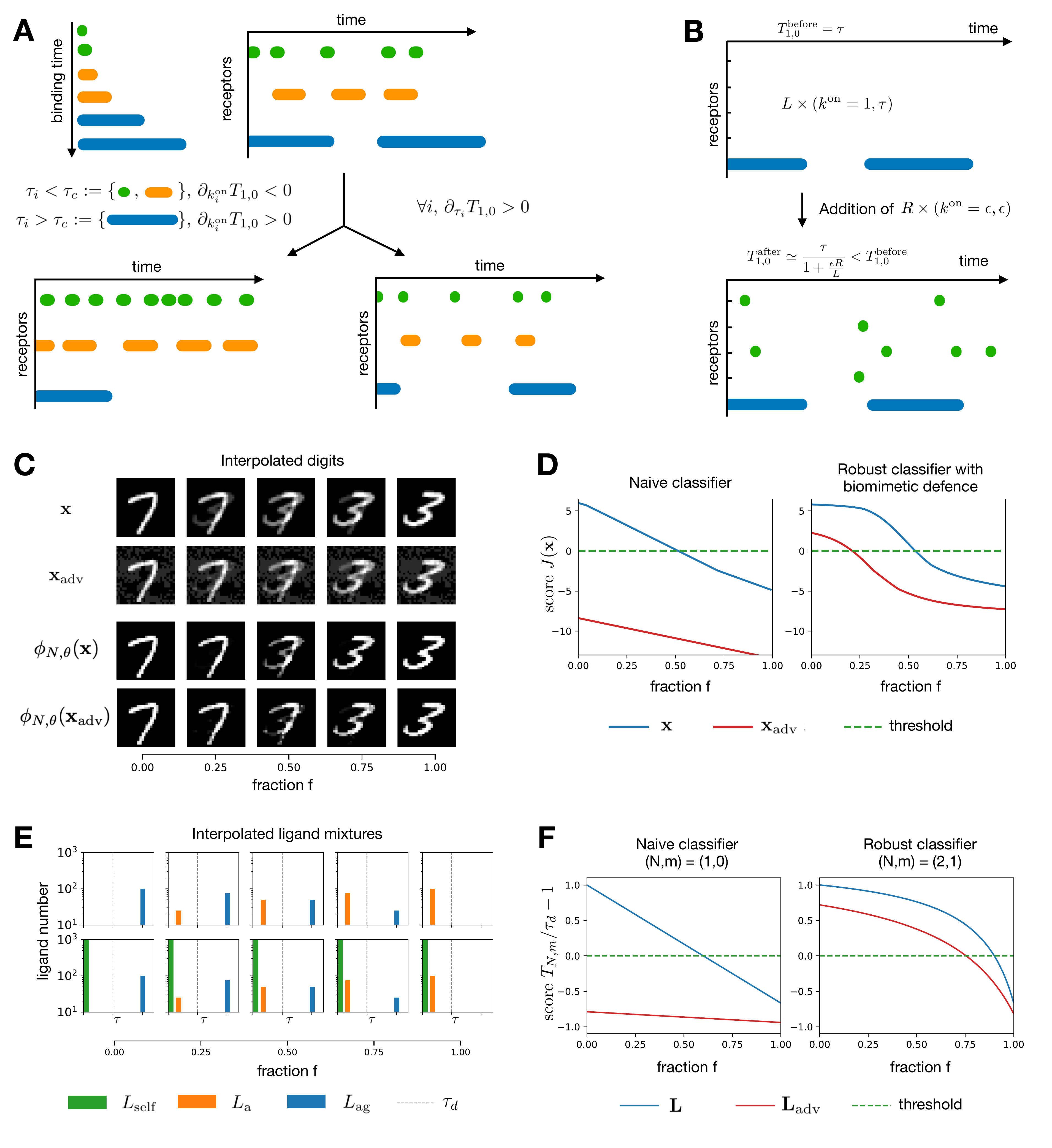}
\caption{ \textbf{Schematics of FGSM applied to immune recognition.} (A) We compute how to lower the response for the receptor occupancy through a given period of time by changing $k^\text{on}_i$ and $\tau_i$. Bottom left: increasing $k^\text{on}_i$ for ligands with $\tau_i < \tau_d$ and decreasing $k^\text{on}_i$ for ligands with $\tau_i > \tau_d$ reduces the weighted average $T_{1,0}$ (change in frequency of the colored bars). Bottom right: decreasing $\tau_i$ for all ligands decreases $T_{1,0}$ (change in length of the colored bars). (B) Response to non-self ligands is lowered from $T_{1,0}^\text{before}$ to $T_{1,0}^\text{after}$ upon addition of $R$ ligands with small binding time $\epsilon$. (C) Interpolated digits with and without adversarial perturbation along the interpolation axis between $\vec{7}$ ($f=0$) and $\vec{3}$ ($f=1$). Adversarial perturbations are computed via the FGSM with $\epsilon = 0.2$. For the biomimetic defence $\phi(N,\theta)$, we choose $N = 5$ and $\theta = 0.5.$ (D) Scoring function $J(\mathbf{x})$ on pictures of panel C without (left) and with (right) the biomimetic defence. The classification threshold is indicated by the dashed green line at $J = 0$. Samples with $J > 0$ are classified as $7$, otherwise $3$. (E) Interpolated ligand mixtures with and without self ligands along the interpolation axis between agonist ($f=0$) and antagonist ($f=1$). Here, $(L_\text{ag}, \tau_\text{ag}) = (100,6); \, (L_\text{a}, \tau_\text{a}) = (100,1); \, (L_\text{self}, \tau_\text{self}) = (1000,0.1)$ (F) Scoring function on ligand mixtures of panel E for a naive immune classifier $(N,m) = (1,0)$ (left) and a robust immune classifier $(N,m) = (2,1)$ (right). The threshold is indicated by a dashed green line at $T_{N,m} / \tau_d - 1 = 0$. $T_{N,m} / \tau_d - 1 > 0$ corresponds to detection of agonists, below corresponds to no detection. In both digit recognition and ligand discrimination, the naive networks interpolate the score linearly and are sensitive to adversarial perturbations, while the score for robust networks is flatter, closer to the initial samples for longer, thus more resistant to perturbation. 
}
%(Caption on the following page.) 
\label{fig:FGSM}
\end{figure*}

Starting from the computation of $\nabla_x{J}$ with respect to parameters $k^\text{on}_i$ and $\tau_i$, the FGSM perturbation is:
%\begin{eqnarray}
%\partial_{k^\text{on}_i} J &=& J'(T_{1,0})  \frac{(\tau_i-T_{1,0})}{\sum k^\text{on}_i } \label{derivk}\\
%\partial_{\tau_i} J &=&  J'(T_{1,0})  \frac{k^\text{on}_i }{\sum k^{on}_i }\label{derivt}
%\end{eqnarray}
\begin{equation}
\eta = \epsilon \, \text{sgn}
\begin{pmatrix}
\partial_{\tau_i} J \\
\partial_{k^\text{on}_i} J
\end{pmatrix}
= \epsilon \, \text{sgn}(A) \text{sgn}
\begin{pmatrix}
k^\text{on}_i \\
\tau_i-T_{1,0}
\end{pmatrix}, \label{FGSM_param}
\end{equation}
where $A = \frac{J'(T_{1,0})}{\sum k^\text{on}_i} > 0$. Notice in the above expression that since derivatives act on  different parameters, an $\epsilon$ sized-perturbation of a given parameter is expressed in its corresponding unit.  For simplicity we will not explicitly write the conversion factor between units (this is for mathematical convenience and does not impact our results). From the above expression, we find that an equivalent maximum adversarial perturbation is given by three simple rules (Fig.~\ref{fig:FGSM} A).
\begin{itemize}
\setlength\itemsep{0.1em}
\item Decrease all $\tau_i$ by $\epsilon$
\item Decrease $k^\text{on}_i$ by $\epsilon$ for ligands with $\tau_i > T_{1,0}$
\item Increase $k^\text{on}_i$ by $\epsilon$ for ligands with $\tau_i < T_{1,0}$
\end{itemize}

The key relation to adversarial examples from \cite{Goodfellow2014}  comes from considering what happens to the  unbound receptors for which  both $k^\text{on}_i$ and $\tau_i$ are initially $0$. Let us consider a situation with $L$ identical bound ligands with $(k^\text{on}=1$, binding time $\tau$) giving response $T^\text{before}_{1,0}=\tau$ where $\tau$ itself is of order $1$ (i.e. much bigger than the $\epsilon$-sized perturbation on binding time considered in Eq. \ref{FGSM_param} ). The three rules above imply that we are to decrease binding time by $\epsilon$, and that all $R$ previously unbound receptors are now to be bound by ligands with $k^\text{on}=\epsilon$, with small binding time $\epsilon$. We compute the new response to be

\begin{equation}
T^\text{after}_{1,0}=\frac{L(
\tau-\epsilon)+\epsilon R \epsilon }{L+\epsilon R}=\frac{\tau-\epsilon + \frac{ \epsilon R}{L} \epsilon}{1+\frac{\epsilon R}{L}}
\label{anta}
\end{equation}
If there are many receptors compared to initial ligands, and  assuming $\epsilon \ll \tau$, the relative change 
\begin{equation}
\frac{T^\text{after}_{1,0}-T^\text{before}_{1,0}}{T^\text{before}_{1,0}} \simeq -\frac{\frac{\epsilon R}{L}}{1+\frac{\epsilon R}{L}}
\end{equation}
is of order $1$ when $\epsilon R \sim L$, giving a decrease comparable to the original response instead of being of order $\epsilon$ as we would naturally expect from small perturbations to all parameters. Thus, if a detection process is based on thresholding variable $T_{1,0}$, a significant decrease can happen with such perturbation, potentially shutting down response. Biologically, the limit where $\epsilon R$ is big corresponds to a strong antagonistic effect of many weakly bound ligands. Examples can be found in mast cell receptors for immunoglobin: weakly binding ligands have been suggested to impinge a critical kinase thus preventing high affinity ligands to trigger response \cite{Torigoe1998}, a so-called ``dog in the manger" effect. Another example is likely found in detection by NK cells \cite{Das2010}. A similar effect called ``competitive antagonism'' is also observed in olfaction where ligands with strong inverse affinity can impinge action of other ligands \cite{Reddy2018}. One difference in olfaction is that for competitive antagonism, the concentration $C$ is of order $1$ while the affinity $\kappa^{-1}$ is big, conversely, here the concentration $R$ is big while $k^\text{on}$ is low. Since we consider the product of both terms, both situations lead to similar effects, but our focus on a small change of $k^\text{on}$ makes the comparison with machine learning more direct.

%This limit is not unreasonable if we expect many receptors $R$ to be initially unbound. 

\vspace{-10pt}
\subsection{Behaviour across boundaries in sample space and adversarial perturbations}
\vspace{-10pt}

To further illustrate the correspondence, we compare the behaviour of a trained neural network classifying $3$s and $7$s with the adaptive proofreading model $(N,m)=(1,0)$ for more general samples. We build linear interpolations between two samples on either side of the decision boundary for both cases (Fig.~\ref{fig:FGSM} C--F, linear interpolation factor $f$ varying between $0$ and $1$). This interpolation is the most direct way in sample spaces to connect objects in two different categories. The neural network classifies linearly interpolated digits, while the adaptive proofreading model classifies gradually changing ligand distributions. 

We plot the output of the neural network $x$ just before taking the sigmoid function $\sigma$ defined in  Fig.~\ref{fig:networks} C and similarly, we plot $T_{N,m}/\tau_d-1$ for adaptive proofreading models. In both cases the decision is thus based on the sign of the considered quantity. In the absence of adversarial/antagonistic perturbations, for both cases, we see that the score of the system almost linearly interpolates between values on either side of the classification boundary (top panel of Fig.~\ref{fig:FGSM} D, F, blue curves). However, in the presence of adversarial/antagonistic perturbations, the entire response is shifted way below the decision boundary (top panel of Fig.~\ref{fig:FGSM} D, F, red curves), so that in particular the initial samples at $f = 0$ (image of 7 or ligand distribution above threshold) are strongly misclassified.

Goodfellow et al. \cite{Goodfellow2014} proposed the linearity hypothesis as an explanation for this adversarial effect: adding $\eta=\epsilon \text{ sgn}\left( \nabla_x J \right)$ to the image leads to a significant perturbation on the scoring function $J$ of order $\epsilon d$, with $d$  the usually high dimensionality of the input space. Thus many weakly lit up background pixels in the initial image can conspire to fool the classifier, explaining the significant shift in the scoring function in Fig.~\ref{fig:FGSM} D top panel. This is consistent with the linearity  we observe on the interpolation line even without adversarial perturbations. A more quantitative explanation based on averaging is given in \cite{Tsipras2018} on a toy-model, that we reproduce below to further articulate the analogy: after defining a label $y \in \{-1,+1\}$, a fixed probability $p$ and a constant $\eta$, one can create a $(d+1)$ dimensional feature vector $x$.

\begin{equation}
\begin{array}{r}
y \in \{ - 1 , + 1 \} , \quad x _ { 1 } \sim \left\{ \begin{array} { l l l } { + y , } & { \mathrm { w.p. } } & {p} \\ { - y , } & { \mathrm { w.p. } } & {1 - p} \end{array} \right. \\ \quad x _ { 2 } , \ldots , x _ { d + 1 } \in \mathcal { N } ( \eta y , 1 )
\end{array}\label{MetaF}
\end{equation}

%\begin{equation}
%y \stackrel { u . a \cdot r } { \sim } \{ - 1 , + 1 \} , \quad x _ { 1 } = \left\{ \begin{array} { l l l } { + y , } & { \mathrm { w.p. } } & {p} \\ { - y , } & { \mathrm { w.p. } } & {1 - p} \end{array} \right., \quad x _ { 2 } , \ldots , x _ { d + 1 } \stackrel { i . i . d } { \sim } \mathcal { N } ( \eta y , 1 )
%\end{equation}

From this, Tsipras et al. build a 100\% accurate classifier in the limit of $d \rightarrow \infty$ by averaging out the weakly correlated features $x_2, \dots x_d$, which gives the score $f_\text{avg}=\mathcal{N}(\eta y,\frac{1}{d})$. Taking the sign of $f_\text{avg}$ will coincide with the label $y$ with $99 \%$ confidence for $\eta \geq 3/\sqrt{d}$.  But such classification can be easily fooled by adding a small perturbation $\epsilon = -2 \eta y$ to every component of the features, since it will shift the average by the same quantity $-2 \eta y$, which can still be small if we take $\eta = O(1/\sqrt{d})$  \cite{Tsipras2018}.
 
We observe a  very similar effect in the simplest adaptive proofreading model. The strong shift of the average $T_{1,0}$ in Eq.~\ref{anta} is due to weakly bound receptors $\epsilon R$, which play the same role as the weak features (components $ x _ { 2 } , \ldots , x _ { d + 1}$ above), hiding the ground truth given by ligands of binding time $\tau$ (equivalent to $x_1$ above) to fool the classifier. We also see a similar linearity on the interpolation in  Fig.~\ref{fig:FGSM} F top panel. There is thus a direct intuitive correspondence between adversarial examples in machine learning and many weakly bound ligands. In both cases, the change of scoring function (and corresponding misclassification) can be large despite the small amplitude $\epsilon$ of the perturbation. Once this perturbation is added, the system in Fig.~\ref{fig:FGSM} still interpolates between the two scores in a linear way, but with a strong shift due to the added perturbation.

\vspace{-10pt}
\subsection{Biomimetic defence for digit classification inspired by adaptive sorting}
\vspace{-10pt}
Kinetic proofreading, famously known as the error-correcting mechanism in DNA replication \cite{Hopfield1974,Ninio1975}, has been proposed as a mechanism for ligand discrimination \cite{McKeithan1995}. In the adaptive proofreading models we are studying here, kinetic proofreading allows the encoding of distinct $\tau$ dependencies in the activation/repression branches \cite{Lalanne2013}. The primary effect of kinetic proofreading is to nonlinearly decrease the relative weight of weakly bound ligands with small binding times, thus ensuring defence against antagonism by weakly bound ligands. Inspired by this idea, we implement a simple defense for digit classification. Before feeding a picture to the neural network, we transform individual pixel values $x_i$ of image $\mathbf{x}$ with a Hill function as

\begin{equation}
x_i \leftarrow \phi_{N,\theta}(x_i) = \frac{x_i^N}{x_i^N+\theta^N}, \label{biomimetic}
\end{equation}

where $N$ (coefficient inspired by kinetic proofreading) and $\theta \in [0,1]$ are parameters we choose. Similarly to the defence of adaptive proofreading where ligands with small $\tau$ are filtered out, this transformation squashes greyish pixels with values below threshold $\theta$ to black pixels, see Fig.~\ref{fig:FGSM} C bottom panels.

In Fig.~\ref{fig:FGSM} D, bottom panel, we show the improved robustness of the neural network armed with this defence. Here, the adversarial perturbation is filtered out efficiently. Strikingly, with or without adversarial perturbation, the score now behaves nonlinearly along the interpolation line in sample space: it stays flatter over a broad range of $f$ until suddenly crossing the boundary when the digit switches identity (even for a human observer) at $f=0.5$. Similarly, for adaptive sorting with $(N,m)=(2,1)$, antagonism is removed, and the score exhibits the same behaviour of flatness followed by a sudden decrease on the interpolation line. Thus, similar defence displays similar robust behaviour of the score in sample space.

\vspace{-10pt}
\subsection{Gradient dynamics identify two different regimes}
\vspace{-10pt}
The dynamics of the score along a trajectory in sample space can thus vary a lot as a function of the model considered. This motivates a more general study of a worst-case scenario, i.e. gradient descent towards the decision boundary for different models. Krotov and Hopfield studied a similar problem for an MNIST digit classifier, encoded with generalized Rectified polynomials of variable degrees $n$ \cite{Krotov2017} (reminiscent of the iterative FGSM introduced in \cite{Kurakin2016}). The general idea is to find out how to most efficiently reach the decision boundary, and how this depends on the architecture of the decision algorithm. Krotov and Hopfield identified a qualitative change with increasing $n$, accompanied by a better resistance to adversarial perturbations \cite{Krotov2016,Krotov2017}.

We consider the same problem for adaptive proofreading models, and study the potential-derived dynamics of binding times for a ligand mixture with identical $k_\text{on}$ when following the gradient of $T_{N,m}$ (akin to a potential in physics). The adversarial goal is to fool the classifier with a minimal change in a given example (or in biological terms, how to best antagonize it). We iteratively change the binding time of non-agonist ligands $\tau < \tau_d$ to
\begin{equation}
\tau \leftarrow \tau-\epsilon \frac{\partial T_{N,m}}{\partial \tau} 
\label{potential}
\end{equation}
while keeping the distribution of agonist ligands with $\tau > \tau_d$ constant. In the immune context, these dynamics can be thought of as a foreign agent selected by evolution to antagonize the immune system. Some biological constraints will force ligands to stay above threshold, so the only possible evolutionary strategy is to mutate and generate antagonists ligands to mask its non-self part. Such antagonistic phenomena have been proposed as a mechanism for HIV escape \cite{Klenerman1994,Meier1995} and associated vaccine failure \cite{Kent1997}. Similar mechanisms might also be implicated in the process of tumour immunoediting \cite{Schumacher2015}.

\begin{figure*}[!thb]
\centering
\includegraphics[width=\linewidth]{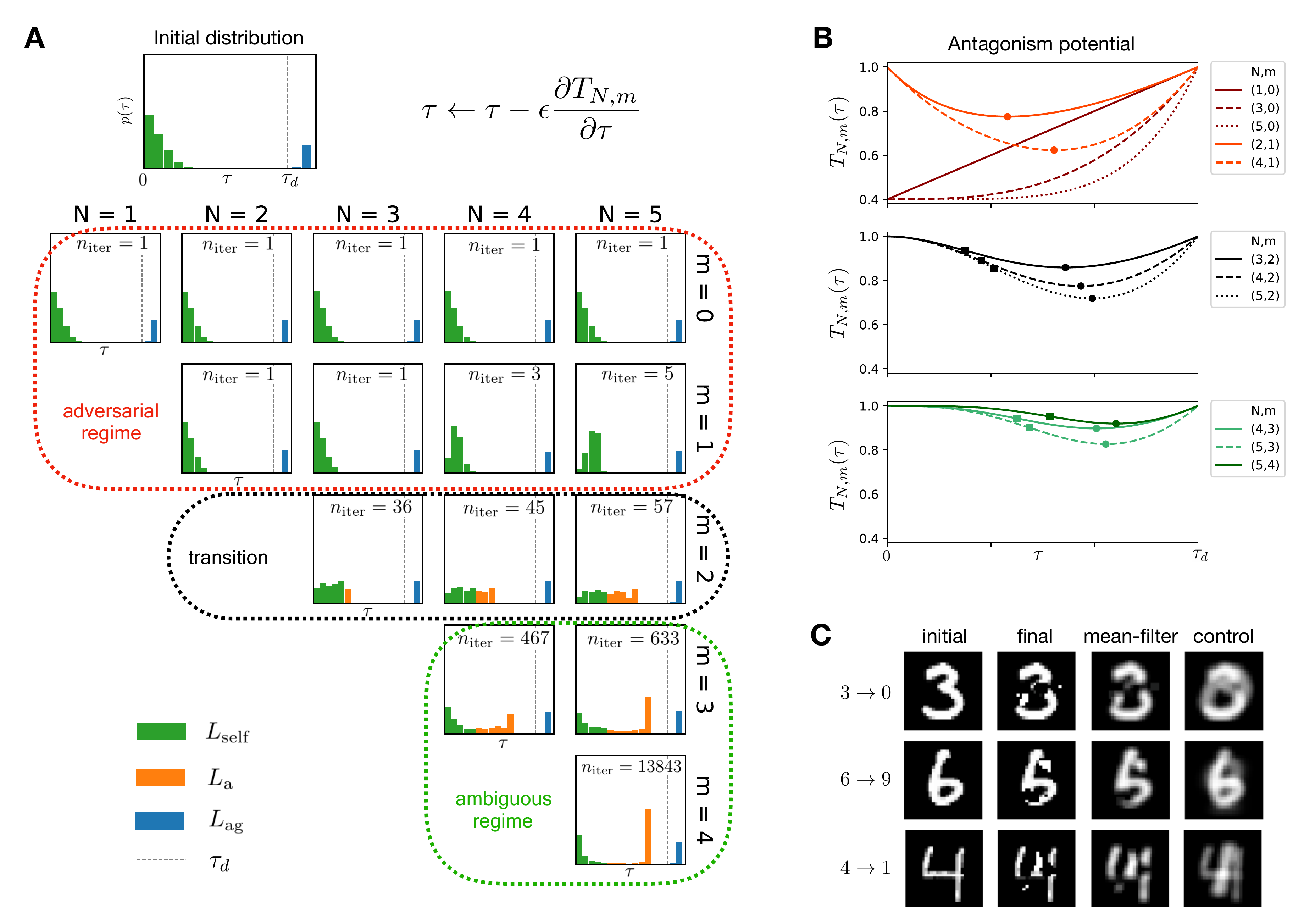}
\caption{\textbf{Characterization of the decision boundary following gradient descent dynamics.} (A) Ligand distribution at the decision boundary by applying iterative gradient descent (top right of the panel) to an initial distribution (top left). For various cases $(N,m)$ we change the binding time of self ligands along the steepest gradient until reaching the decision boundary. $n_\text{iter}$ indicates the number of iterations needed to reach the decision boundary. We identify the adversarial regime (red), the ambiguous regime (green) and a transition (black) depending on $m$. (B) $T_{N,m}$ for mixtures of ligands at $\tau_d$ and ligands at $\tau$, as a function of $\tau$ for various $(N,m)$. Antagonism strength is maximal when $T_{N,m}$ is minimal. Minima and inflexion points are indicated with a circle and square. (C) Few-pixel attack as a way of circumventing proofreading or local contrast defence, while creating ambiguous digits. We add a 3x3 mean-filter to demonstrate the ambiguity of digits at the decision boundary. The control image is the mean filtered initial digit combined with the locally contrasted average target digit. Note that also the control is lacking a clear ground truth.} 
\label{MTL}
\end{figure*}

From a given ligand mixture with few ligands above threshold and many ligands below thresholds, we follow the dynamics of Eq.~\ref{potential}, and display the ligand distribution at the decision boundary for different values of $N,m$ as well as the number of steps to reach the decision boundary in the descent defined by Eq.~\ref{potential} (Fig.~\ref{MTL}, see also Fig.~S\ref{fig:MTL} for another example with a visual interpretation). We observe two qualitatively different dynamics. For $m <2$, we observe strong adversarial effects, as the boundary is almost immediately reached and the ligand distribution barely changes. As $m$ increases, in Fig.~\ref{MTL} A the ligands in the distribution concentrate around one peak. For $m=2$, a qualitative change occurs: the ligands suddenly spread over a broad range of binding times and the number of iterations in the gradient dynamics to reach the boundary drastically increases. For $m>2$, the ligand distribution becomes bimodal, and the ligands close to $\tau = 0$ barely change, while a subpopulation of ligands peaks closer to the boundary. Consistent with this, the number of $\epsilon$-sized steps to reach the boundary is 3 to 4 orders of magnitude higher for $m>2$ as for $m<2$.  

\vspace{-10pt}
\subsection{Qualitative change in dynamics is due to a critical point for the gradient}
\vspace{-10pt}

The qualitative change of behaviour observed at $m=2$ can be understood by studying the contribution to the potential $T_{N,m}$ of ligands with very small binding times $\tau_\epsilon \sim 0$. Assuming without loss of generality that only two types of ligands are present (agonists $\tau_\text{ag}>\tau_d$ and spurious $\tau_\text{spurious}=\tau_\epsilon$), an expansion in $\tau_{\epsilon}$ gives, up to a constant, $T_{N,m} \propto - \tau_{\epsilon}^{m}$ for small $\tau_\epsilon$ (see  Fig.~\ref{MTL} B for a representation of this potential and Appendix S3 for this calculation). In particular, for $0<m<1$, $\frac{\partial T_{N,m}}{\partial \tau_\epsilon} \propto -\tau_\epsilon^{m-1}$ diverges as $\tau_\epsilon \rightarrow 0$. This corresponds to a steep gradient of $ T_{N,m}$ so that the system quickly reaches the boundary in this direction. The ligands close to $\tau_\epsilon \sim 0$ then quickly localize close to the minimum of this potential (unimodal distribution of ligand for small $m$ on Fig.~\ref{MTL} A--B). 

The potential close to $\tau_\epsilon \sim 0$ flattens for $1<m<2$, but it is only at $m=2$ that a critical point for the \textit{gradient} (i.e. characterized by $ \partial^2 T_{N,m}/(\partial \tau_\epsilon)^2 =0$) appears at $\tau_\epsilon=0$. This qualitatively modifies the dynamics defined by Eq.~\ref{potential}. For $m \geq 2$, due to the new local flatness of this gradient, ligands at $\tau=0$, the dynamical critical point of Eq.~\ref{potential}, are pinned by the dynamics. By continuity, dynamics of the ligands slightly above $\tau_\epsilon=0$ are critically slowed down, making it much more difficult for them to reach the boundary. This explains both the sudden broadening of the ligand distribution, and the associated increase in the number of steps to reach the decision boundary.  Conversely, an inflexion point (square) appears in between the minimum (circle) and $\tau_\epsilon = 0$ (Fig.~\ref{MTL} B). Ligands close to the inflexion point separate and move more quickly towards the minimum of potential, explaining the bimodality at the boundary (if we were continue the dynamics past the boundary, all ligands with non-zero binding times would collapse to the minimum of the potential). For both larger $N$ and larger $m$ we obtain flatter potentials, and  a larger number of iterations. In Appendix S4, we further describe the consequence of adding proofreading steps on the position of the boundary itself, using another concept of machine learning called ``boundary tilting'' \cite{Tanay2016} (Fig.~S\ref{fig:tilt} and Table S1). %, allowing to better categorize perturbations in terms of local gradients.

%Noteworthy, increasing $N$ for fixed $m$ speeds up convergence to the boundary because of a stiffer gradient, indicating that more proofreading steps is not favorable for immune detection of such changing agents. NO LONGER TRUE ??

\vspace{-10pt}
\subsection{Categorization of attacks}
\vspace{-10pt}

The transition at $m =2$ is strongly reminiscent of the transition observed by Krotov and Hopfield in their study of gradient dynamics similar to Eq.~\ref{potential} \cite{Krotov2017}. In both our works, we see that there are (at least) two kinds of attacks that can bring samples to the decision boundary. The FGSM corresponds to small perturbations to the input in terms of $L_\infty$ norm leading to modifications of many background pixels in \cite{Krotov2017} or many weakly bound ligands for the adaptive proofreading case, also similar to the meaningless changes in $x_2, \dots x_d$ described above in Eq.~\ref{MetaF} \cite{Tsipras2018}.

Defence against the FGSM perturbation is implemented through a higher degree $n$ of the rectified polynomials in \cite{Krotov2017}, while in adaptive proofreading, this is done through critical slowing down of the dynamics of Eq.~\ref{potential} for $m>2$. The latter models are nevertheless sensitive to another kind of attack with many fewer perturbations of the inputs but with bigger magnitude. This corresponds to digits at the boundary where few well-chosen pixels are turned on in \cite{Krotov2017}. For adaptive proofreading models this leads to the ligand distribution becoming bimodal at the decision boundary. Three important features are noteworthy. First, the latter perturbations are difficult to find through gradient descent (as illustrated by the many steps to reach the boundary in Fig.~\ref{MTL}A). Second, the perturbations appear to be meaningful: they correspond to interpretable features and interfere with the original sample. These perturbations make it difficult or even impossible to recover the ground truth by inspecting the sample at the decision boundary. Digits at the boundary for \cite{Krotov2017} appear indeed ambiguous to a human observer, and ligand distribution peaking just below threshold are potentially misinterpreted biologically due to inherent noise. This has actually been observed experimentally in T cells, where strong antagonists are also weak agonists \cite{Altan-Bonnet2005,Francois2013}, meaning that T cells do not take reliable decisions in this regime. Lastly, it has been observed in machine learning that memory capacity considerably increases for high $n$ in \cite{Krotov2016}, due to the local flattening of the landscape close to memories (ensuring that random fluctuations do not change memory recovery). A similar effect in our case is observed: the antagonism potential is flattened out with increasing $N,m$ so that any spurious antagonism becomes at the same time less important and lies closer to the decision boundary.

\vspace{-10pt}
\subsection{Biomimetic defenses against few-pixel attacks}
\vspace{-10pt}

It is then worth testing the sensitivity  to localized stronger attacks of digit classifiers, helped again with biomimetic defences. The natural analogy is to implement attacks based on strong modification of few pixels \cite{Su2017}. 

For this problem, we choose to implement a two-tier biomimetic defence:  we implement first the transformation defined in Eq.~\ref{biomimetic}, that will remove influence of the FGSM types of perturbations by flattening the local landscape as in Fig.~\ref{fig:FGSM} D. In addition, we choose to add a second layer of defence where we simply average out locally pixel values. This can be interpreted biologically as a process of receptor clustering or time-averaging. Time-averaging has been shown to be necessary in a stochastic version of adaptive proofreading \cite{Francois2013,Lalanne2013}, where temporal intrinsic noise would otherwise make the system cross the boundary back and forth endlessly. In the machine learning context, local averaging has been recently proposed as a way to defend against few pixel attacks \cite{Xie2018}, which thus can be considered as the analogous of defending against biochemical noise.

We then train multiple classifiers between different pairs of handwritten digits. Following the approach of the ``one pixel'' attack \cite{Su2017}, we consider digits classified in presence of this two-tier defence, then sequentially fully turn pixels on or off ranked by their impact on the scoring function, until we reach the decision boundary. Details on the procedure are described in Appendix S5. A good defence would manifest itself similarly to the Krotov-Hopfield case  \cite{Krotov2017}, where no recognizable (or ambiguous) digits are observed at the boundary.

Representative results of such few-pixel attacks with biomimetic defences are illustrated in Fig.~\ref{MTL} C. The ``final" column shows the misclassified digits after the attack and the ``mean-filter" column shows the local average of the ``final" digits for further comparison, with other examples shown in Fig.~S\ref{fig:few-pixel-attack} and details on the behaviour of scoring functions in Fig.~S\ref{fig:score-few-pixel-attack}. Clearly the attacked samples at the boundaries hide the ground truth of the initial digit, and as such can not be considered as typical adversarial perturbations. Samples at the boundary are out-of-distribution but preserve structure comparable to written characters (e.g. attacks from $0$ to $1$ typically look like a Greek $\phi$, see Fig.~S\ref{fig:few-pixel-attack}). This makes them impossible to classify as Arabic digits even for a human observer. This is consistent with the ambiguous digits observed for big $n$ by Krotov and Hopfield \cite{Krotov2017}. In other cases, samples at the boundary between two digits actually look like a third digit: for instance, we see that the sample at the boundary between a 6 and an 9 looks like a 5 (or a Japanese \begin{CJK*}{UTF8}{min}ち\end{CJK*}). This observation is consistent with previous work attempting to interpolate in latent space between digits \cite{Berthelot2018}, where at the boundary a third digit corresponding to another category may appear. We also compare in Fig.~\ref{MTL} C the sample seen by the classifier at the boundary after the biomimetic defences with a ``control'' corresponding to the average between the initial digit and the target of the attack (corresponding to the interpolation factor $f=0.5$ in Fig.~\ref{fig:FGSM} C--D). It is then quite clear that the sample generated by the attack is rather close to this control boundary image. This, combined with the fact that samples at the boundary still look like printed characters without clear ground truth indicate that the few pixel attacks implemented here actually select for meaningful features. The existence of meaningful features in the direction of the gradient have been identified as a characteristic of networks robust to adversarial perturbation \cite{Tsipras2018} similar to results of \cite{Krotov2017} and our observation for adaptive proofreading models above. 

% Samples at the boundary superficially look like printed Japanese ``kanas" or Greek characters

\vspace{-10pt}
\section{DISCUSSION}
\vspace{-10pt}

Complex systems (\textit{in vivo} or \textit{in silico}) integrate sophisticated decision making processes. Our work illustrates common features between neural networks and a general class of adaptive proofreading models, especially with regards to mechanisms of defence against targeted attacks. Parallels can be drawn between these past approaches, since the models of adaptive proofreading presented here were first generated with \textit{in silico} evolution aiming at designing immune classifiers \cite{Lalanne2013}. Strong antagonism naturally appeared in the simplest simulations, and required modification of objective functions very similar to adversarial training \cite{Goodfellow2014}.

Through our analogy with adaptive proofreading, we are able to identify the presence of a critical point in the gradient of response as the crucial mediator of robust adversarial defense. This critical point emerges due to kinetic proofreading for cellular decision network, and essentially removes the spurious adversarial directions. Another layer of defence can be added with local averaging. This is in line with current research on adversarial robustness in machine learning, showing that robust networks exhibit a flat loss landscape near each training sample \cite{Moosavi-Dezfooli2018b}. Other current explorations include new biomimetic learning algorithms, giving rise to prototype-like classification \cite{Krotov2019}. Adversarial defence strategies, including non-local computation and nonlinearities in the neural network are also currently under study \cite{Xie2018}. The mathematical origin of the effectiveness of those defences is not yet entirely clear, and identification of critical points in the gradient might provide theoretical insights into it.

% If the gradient is flat close to a sample far from the boundary, nonzero at the boundary, then flat close to another sample, an inflexion point is expected via Rolle's theorem
More precisely, an interesting by-product of local flatness, where both the gradient and second derivative of the score are equal to zero, is the appearance of an inflexion point in the score, and thus a region of maximal gradient. This is visible in Fig.~\ref{fig:FGSM} D, F: while the score of non-robust classifiers is linear when moving towards the decision boundary, the scoring function of classifiers resistant to adversarial perturbations is flat at $f=0$ and only significantly changes when the input becomes ambiguous near the inflexion point. The reason why this is important in general is that a combination of local flatness and an inflexion point is bound to strongly influence any gradient descent dynamics. For instance, for adaptive proofreading models, the ligand distribution following the dynamics of Eq.~\ref{potential} changes from unimodal to bimodal at the boundary, creating ambiguous samples. For a robust classifier, such samples are thus expected to appear close to the decision boundary since they coincide with the larger gradients of the scoring function. As such they could correspond to meaningful features (contrasting the adversarial perturbations), as we show in Fig.~\ref{MTL} C with our digit classifier with biomimetic defence. Examples in image classification might include the meaningful adversarial transformations between samples found in \cite{Tsipras2018}  or the perturbed animal pictures fooling humans \cite{Elsayed2018} with chimeric images that combine different animal parts (such as spider and snake), leading to ambiguous classifications. Similar properties have been observed experimentally for ambiguous samples in immune recognition: maximally antagonizing ligands have a binding time just below the decision threshold \cite{Altan-Bonnet2005}. We interpret this property as a consequence of the flat landscape far from the decision threshold leading to a steeper gradient close to it \cite{Francois2013,Francois2016b}.  

% to perfectly align samples along the $\tau$ direction
%. As shown in the Appendix S4 for the 3 vs 7 problem, the linear transformation to the first two PCA modes already aligns digits along the decision axis

We used machine learning classification and implemented biomimetic defence by relying on a single direction, since that is what emerges in the most simple version of adaptive proofreading models that we considered here.
%A possible caveat of our analogy is that a clear decision axis in the $\tau$ direction can be defined explicitly in adaptive proofreading models - even though it does require the adaptive proofreading architecture to remove the ligand quantity dependence.
In general, however, the space of inputs in machine learning is much more complex, and there are more than two categories, even in digit classification. One possible solution is to break down multilabel classification into a set of binary classification problems, but this might not always be appropriate. Instead, the algorithm effectively has to learn representations, such as pixel statistics and spatial correlations in images \cite{Krizhevsky2012}. With a nonlinear transformation to a low-dimensional manifold description, one could  still combine information on a global level in ways similar to parameter $\tau$. The theory presented here could then apply once the mapping of the data from the full-dimensional space to such latent space is discovered. 

Case-in-point, Tsipras et al. proposed a distinction in machine learning between a robust, but probabilistic feature ($x_1$ in Eq. \ref{MetaF}) and weakly correlated features ($x_2, \dots x_d$ in Eq. \ref{MetaF})  \cite{Tsipras2018}, both defining a single direction in latent space. They then observed a robustness-accuracy trade-off due to the fact that an extremely accurate classifier would mostly use a distribution of many weakly correlated features (instead of the robust -- but randomized -- feature) to improve accuracy. The weight to put in the decision on either feature (robust or weak) would depend on the training. Our work shows the natural connection between weak features in this theory and weak ligands in the biological models (see discussion below  Eq. \ref{MetaF}). In the biological context, the standard situation is that all ligands are treated equally. Then one can show mathematically that for such networks performing quality sensing irrespective of quantity, antagonism necessarily ensues \cite{Francois2016b}, as further identified here using the FGSM transformation. This latter result can be reformulated in terms of machine learning \cite{Tsipras2018} in the following compact way: perfectly robust  classification (i.e. with no antagonism) is impossible in biology if all receptors are equivalent. But biology also provides evidence that robustness can nevertheless be improved by applying local nonlinear transformation such as the biomimetic defence of Eq. \ref{biomimetic}. Elaborating on the distinction between robust and weak features proposed in  \cite{Tsipras2018}, nonlinear transformations should specifically target weak correlated features. Explorations of generalized nonlinear transformations in image feature space \cite{Krotov2016,Krotov2017} might lead to further insights into the possible nonlinear transformations defending against adversarial perturbations. We learn in particular from biology that the major effect of nonlinearity is to change the position of maximally adversarial perturbations in sample space. Perfect robustness might be impossible in general, yet similarly to cellular decision-making the most effective perturbations may shift from a pile of apparently unstructured features for naive classifiers to a combination of meaningful features for robust classifiers, giving ambiguous patterns at the decision boundary (allowing to further distinguish between ambiguous and adversarial perturbations).

%This is precisely due to the necessary presence of a significant gradient in the direction of the decision-making. Quenching influences of weak ligands with nonlinearities change the binding time of maximally antagonizing ligands. 

%It was also shown mathematically that for the classification problem of discriminating ligand quality irrespective of their quantity, one always gets antagonism close to the boundary \cite{Francois2016b}, but that one can also change the binding time of maximally antagonizing ligands via the nonlinearities in kinetic proofreading. Similar results might be generalizable to machine learning. 

% In immediate follow-up work, we can quantify the robustness-accuracy trade-off in adaptive proofreading models, showing the existence of upper bounds like Theorem 2.1 in Ref. \cite{Tsipras2018}.Such robustness will add an additional constraint to the energy-speed-accuracy trade-off in kinetic proofreading \cite{Murugan2012} and early immune recognition \cite{Cui2018}.

From the biology standpoint, new insights may come from the general study of computational systems built via machine learning. In particular, systematic search and application of adversarial perturbations in both theoretical models and experiments might reveal new biology. For instance, our study of Fig.~\ref{MTL}, inspired by gradient descent in machine learning \cite{Krotov2017}, establishes that cellular decision-makers exist in two qualitatively distinct regimes. The difference between these regimes are geometric by nature through the presence or absence of a dynamical critical point in the gradient. The case $m<2$ with a steep gradient could be more relevant in signalling contexts to separate mixtures of inputs, so that every weak perturbation \textit{should} be detected \cite{Carballo2018}. For olfaction it has been suggested that strong antagonism allows for a rescaling of the distribution of typical odor molecules, ensuring a broad range of detection irrespective of the quantity of molecules presented \cite{Reddy2018}. The case $m\geq2$ is much more resistant to adversarial perturbations, and could be most relevant in an immune context where T cells filter out antagonistic perturbations. This might be relevant for the pathology of HIV infections \cite{Klenerman1994,Meier1995,Kent1997} or, more generally, could provide explanations on the diversity of altered peptide ligands \cite{Unanue2011}. We also expect similar classification problems to occur at the population-level, e.g. when T cells interact with each other to refine individual immune decision-making \cite{Butler2013,Voisinne2015}. Interestingly, there might be there a trade-off between resistance to such perturbations (in particular to self antagonism, pushing towards higher $m$ in our model) and the process of thymic selection which relies on the fact that there should be sensitivity to some self ligands \cite{Mandl2013} (pushing towards lower $m$ in our model) .

Our correspondence could also be useful for the theoretical modelling and understanding of cancer immunotherapy \cite{Snyder2014}. So-called neoantigens corresponding to mutated ligands are produced by tumours. It has been observed that in the presence of low-fitness neoantigens, the blocking of negative signals on T cells (via checkpoint inhibitor blockade) increases success of therapy \cite{Luksza2017}. This suggests that those neoantigens are ambiguous ligands: weak agonists acting in the antagonistic regime. Without treatment, negative signals prevent their detection (corresponding to an adversarial attack), but upon checkpoint inhibitor blockade those ligands are suddenly visible to the immune system, which can now eliminate the tumour. Importantly, differential responses are present depending on the type of cancer,  environmental factors and tumour microenvironment \cite{Schumacher2015}. This corresponds to different background ligand distributions in our framework, and one can envision that cancer cells adapt their corresponding adversarial strategies to escape the immune system. Understanding and categorizing possible adversarial attacks might thus be important to predicting the success of personalized immunotherapy \cite{Sahin2018}.

We have connected  machine learning algorithms to models of cellular decision-making, and in particular their defence strategies against adversarial attacks. More defences against adversarial examples might be found in the real world, for instance in biofilm-forming in bacteria \cite{Yan2017}, in size estimation of animals \cite{Laan2018}, or might be needed for proper detection of physical 3D objects \cite{Athalye2018} and road signs \cite{Eykholt2018}. Understanding the whole range of possible antagonistic perturbations may also prove crucial for describing immune defects, including immune escape of cancer cells. It is thus important to further clarify possible scenarios for fooling classification systems in both cell biology and machine learning.

\vspace{-10pt}
\section{ACKNOWLEDGEMENTS}
\vspace{-10pt}
We thank Joelle Pineau and members of the Fran\c cois group for useful discussions. We thank three anonymous reviewers for comments and suggestions. P.F. is supported by a Simons Investigator in Mathematical Modelling of Living Systems award, an Integrated Quantitative Biology Initiative award and a Natural Sciences and Engineering Research Council award (Discovery Grant). T.J.R. receives funding from the Centre for Applied Mathematics in Bioscience and Medicine (graduate award), McGill Physics (Schulich award),  and the Fonds de Recherche du Québec - Nature et Technologies (graduate award). E.B. acknowledges support from the Samsung Advanced Institute of Technology and the Fonds de Recherche du Québec – Nature et Technologies (graduate award).

\bibliography{My_Collection}

\onecolumngrid

%Supplementary material section
\setcounter{equation}{0}%Reset equation number for Appendix
\setcounter{figure}{0}%Reset figure number for Appendix
\setcounter{table}{0}%Reset table number for Appendix

% \captionsetup{labelformat=empty}
% \preto{\caption}{\captiondelim{ }} % change \captiondelim

\newpage

\vspace{-10pt}
\section{Appendix}
\vspace{-10pt}
% S1 Appendix
\subsection{Appendix S1: Mathematical details of the adaptive proofreading models}
%\label{supp:adaptivesortingmodel}
\vspace{-10pt}
Appendix S1 contains more details on the derivation of adaptive proofreading models (section \textit{Biochemical kinetics}), referred to in section \textbf{Adaptive proofreading for cellular decision-making} in the main text. We also give the parameters and equations that are used to draw Fig.~\ref{fig:networks} B in the main text (section \textit{Parameters for Fig.~\ref{fig:networks} B}).

% S1 Appendix
\subsubsection{Biochemical kinetics}
%\label{supp:adaptivesortingmodel}
\vspace{-10pt}

The kinetics for the biochemical network in Fig.~\ref{fig:networks} B in the simplest form ($(N,m) = (2,1)$) are given by 

\begin{align}
&\dot C_1 = k^\text{on} R L - (\phi K + \tau^{-1}) C_1 \nonumber \\
&\dot C_2 = \phi K C_1 - \tau^{-1} C_2 \label{eq:kinetics} \\
&\dot K = \beta (K_T - K) - \alpha C_1 K. \nonumber
\end{align}

Here, we assume the T cell has $R$ receptors to which $L$ ligands are bound to form ligand-receptor complexes $C_1$ and $C_2$. The parameters $k^\text{on}$ and $\tau^{-1}$ denote ligand-specific rates, which correspond to an average number of events happening per second (mean of a Poisson-distributed variable). $\phi$ is the phosphorylation rate for the reaction $C_1 \to C_2$ (activation branch), which is activated by variable $K$, and which we will call a generic kinase. $K$ itself is inhibited by $C_1$ (repression branch) with rate $\alpha$. $K_T$ here is the total number of kinase, and $K_T - K$ the number of inactive kinase. This kinase is shared between all receptors and assumed to diffuse freely and rapidly, so that since $K$ is inactivated by $C_1$, (in)activity of $K$ is a measure of the total number of receptors bound. Lastly, $\beta$ is the activation rate of $K$. In the steady state, we can solve exactly for $C_2$ and find

\begin{equation}
C_2 = \phi K C_1 \tau =  \frac{L \tau}{\beta / \alpha + L} \simeq \frac{L \tau}{L} = \tau. 
\label{eq:steady_state}
\end{equation}

Here $K = \frac{K_T \beta / \alpha}{\beta / \alpha + C_1}$, and as long as $L \gg \beta / \alpha$ the first-order approximation is exact and the ligand dependence in nominator and denominator cancels. Without loss of generality, we have set $\frac{\phi K_T \beta}{\alpha} = 1$.

When we consider an environment containing two ligand types with binding times $\tau_\text{ag}$ (agonists) and $\tau_\text{a}$ (antagonists) at concentrations $L_\text{ag}$ and $L_\text{a}$, two types of ligand-receptor complexes can be formed. We call them $C_i$ for agonists and $D_i$ for antagonists. Full equations in the case of $(N,m)=(2,1)$ are given by
\begin{align}
&\dot C_1 = k^\text{on} R L_\text{ag} - (\phi K + \tau^{-1}_\text{ag}) C_1 \nonumber \\
&\dot C_2 = \phi K C_1 - \tau^{-1}_\text{ag} C_2 \\
&\dot D_1 = k^\text{on} R L_\text{a} - (\phi K + \tau^{-1}_\text{a}) D_1 \nonumber \\
&\dot D_2 = \phi K D_1 - \tau^{-1}_\text{a} D_2 \\
&\dot K = \beta (K_T - K) - \alpha (C_1 + D_1) K. \nonumber
\end{align}
where we have assumed that $k^\text{on}$ is equal for both agonist and antagonist ligands. The main difference here is that variable $K$ integrates global information from both ligand complexes, which results in the steady-state in $K = \frac{K_T \beta / \alpha}{\beta / \alpha + C_1 + D_1}$. Moreover, $K$ acts locally on the phosphorylation of both $C_1$ and $D_1$. Finally, the output is given by $T_{2,1} = C_2 + D_2$. 

We can generalize this case by assuming that inhibition of the variable $K$ occurs further downstream a kinetic proofreading cascade, namely at the m-th complex $C_m = L_\text{ag} \tau_\text{ag}^m$ and $D_m = L_\text{a} \tau_\text{a}^m$. The output variable is then given by $T_{N,m} = C_N + D_N$. Fig.~\ref{fig:networks} A shows how information from a single ligand passes through the repression branch (red arrow and box) via $K$ and through the activation branch (green arrow and box) via $C_N$. The global variable $K$ integrates local information as $K = \frac{K_T \beta / \alpha}{\beta / \alpha + C_m + D_m} \propto \left( L_\text{ag} \tau_\text{ag}^m + L_\text{a} \tau_\text{a}^m \right)^{-1}$, and catalyzes the phosphorylation of $C_{N-1} = L_\text{ag} \tau_\text{ag}^{N-1}$ and $D_{N-1} = L_\text{a} \tau_\text{a}^{N-1}$ to final complex $C_N$ and $D_N$ as 

\begin{align}
&\dot{C}_N = K C_{N-1} - \tau_\text{ag}^{-1} C_N \\
&\dot{D}_N = K D_{N-1} - \tau_\text{a}^{-1} D_N.
\end{align}

In the steady-state, the solution for $T_{N,m}$ is then

\begin{equation}
T_{N,m} = C_N + D_N = \frac{L_\text{ag} \tau_\text{ag}^N + L_\text{a} \tau_\text{a}^N}{L_\text{ag} \tau_\text{ag}^m + L_\text{a} \tau_\text{a}^m}.
\end{equation}

This expression for two types of ligands with same $k_{on}$ can be clearly generalized to any types of ligands, giving Eq. \ref{generalTnm} in the main text.

\newpage

%% S2 Appendix
\vspace{-10pt}
\subsection{Appendix S2: Materials and methods}
\vspace{-10pt}

In Appendix S2, we give the parameters and equations that are used to draw Fig.~\ref{fig:networks} B in the main text and we give the hyperparameters used for training the neural networks classifying 3s and 7s. We referred to the latter in section \textbf{Neural networks for artificial decision-making} in the main text.

\subsubsection{Parameters for Fig.~\ref{fig:networks} B}
\vspace{-10pt}

The curves on Fig.~\ref{fig:networks} B, left panel, come from the model given by
\begin{equation}
    T_{4,2}(L) = \frac{1}{\tau_d^2} \frac{L \tau^4}{C_* + L \tau^2}
\end{equation}
with parameter values $C_* = \beta / \alpha = 3000, \, \tau_d = 4s$ and $\tau$ as in the legend. The curves on the middle panel of Fig.~\ref{fig:networks} B come from
\begin{equation}
    T_{4,2}(L) = \frac{1}{\tau_d^2} \frac{L \tau^4 + L_a \tau_a^4}{C_* + L \tau^2 + L_a \tau_a^2}
\end{equation}
with again $C_* = 3000, \, \tau_d = 4s$ and $\tau = 10s$. For blue "agonists alone", $L_a = 0$ , for orange "+ antagonists" $L_a = 10^4$ and $\tau_a = 3s$, and for green "+ self" $L_a = 10^4$ and $\tau_a = 1s$.

\subsubsection{Hyperparameters for training neural network}
We have chosen our hyperparameters as follows: one hidden layer with four neurons feeding into an output neuron, a random 80/20 training/test split with a 10 percent validation split. The cross-entropy loss function is minimized via stochastic gradient descent in maximal 300 iterations with a batch size of 200 and an adaptive learning rate, initiated at 0.001. The tolerance is $10^{-4}$ and the regularization rate is 0.1. Most of these parameters are set to their default value, but we found that the training procedure is largely insensitive to the specific choice of hyperparameters. 

\newpage

%% S3 Appendix
\vspace{-10pt}
\subsection{Appendix S3: Ligand distribution at the decision boundary}
\label{supp:grad_descent}
\vspace{-10pt}

In Appendix S3 we describe in detail the methods used in the gradient dynamics of changing a ligand distribution to the decision boundary (section \textit{Methods}), we provide additional results when adding spatial correlation to the ligand distribution (section \textit{MTL pictures}), and we calculate the leading order in small binding time $\tau_\epsilon$ of the gradient $\frac{dT_{N,m}}{d \tau_\epsilon}$ (section \textit{Behavior for small binding times}). We refer to Appendix S3 in the main text in section \textbf{Gradient dynamics identify two different regimes} and \textbf{Qualitative change in dynamics is due to a critical point for the gradient}, and in Fig.~\ref{fig:FGSM}A.

\subsubsection{Methods}
\vspace{-10pt}

Adaptive proofreading is well-suited to characterize the decision boundary between two classes, because we can work with an analytical description. We want to know how to most efficiently change the binding time of the spurious binding ligand (with small $\tau$) to cause the model to reach the decision boundary. We have taken inspiration from \cite{Krotov2017} and adapted our approach from the iterative FGSM \cite{Kurakin2016}. At first, we sample the binding times $\tau_\text{self}$ for $L_\text{self} = 7000$ self ligands from a half-normal distribution $|\mathcal{N}(0,\frac{1}{3})|$ and $\tau_\text{ag}$ for $L_\text{ag} = 3000 $ agonist ligands from a narrowly peaked normal distribution $|\mathcal{N}(\frac{7}{2},\frac{1}{10})|$ just above $\tau_d=3$. We fix the agonist ligand distribution, the ``signal'' in the immune picture. Next, we bin ligands in $M$ equally spaced bins with center binding time $\tau_b, \, b \in {1,\dots,M}$, and we compute the gradient for bins for which $\tau_b < \tau_d$
\begin{equation}
\frac{\partial T_{N,m}}{\partial \tau_b} = \frac{N \tau_b^{N-1}L_b - m T_{N,m}\tau_b^{m-1} L_b}{\sum_{i=1}^M \tau_i^m L_i}
\end{equation}
where $L_b$ is the number of ligands in the $b^\text{th}$ bin. We subtract this value multiplied by a small number $\epsilon$ from the exact binding times, as in Eq. 6 in the main text, and we compute a new output $T_{N,m}$. We repeat this procedure until $T_{N,m}$ dips just below the response threshold $\tau_d^{N-m}$. We then display the ligand distributions. We bin ligands and compute the gradient in batches to prevent the gradient from becoming negligibly small. If we would compute the gradient for each ligand with an individual binding time, there would be exactly one ligand with that specific binding time, and because the gradient scales with $L$, we would need to go through many more iterations. Decreasing the binsize and step size $\epsilon$ may enhance the resolution, but is not required. We found good results by considering bins with a binsize of 0.2s and $\epsilon = 0.2$. 

\subsubsection{MTL pictures}
\vspace{-10pt}

\begin{figure*}[!thb]
\includegraphics[width=\textwidth]{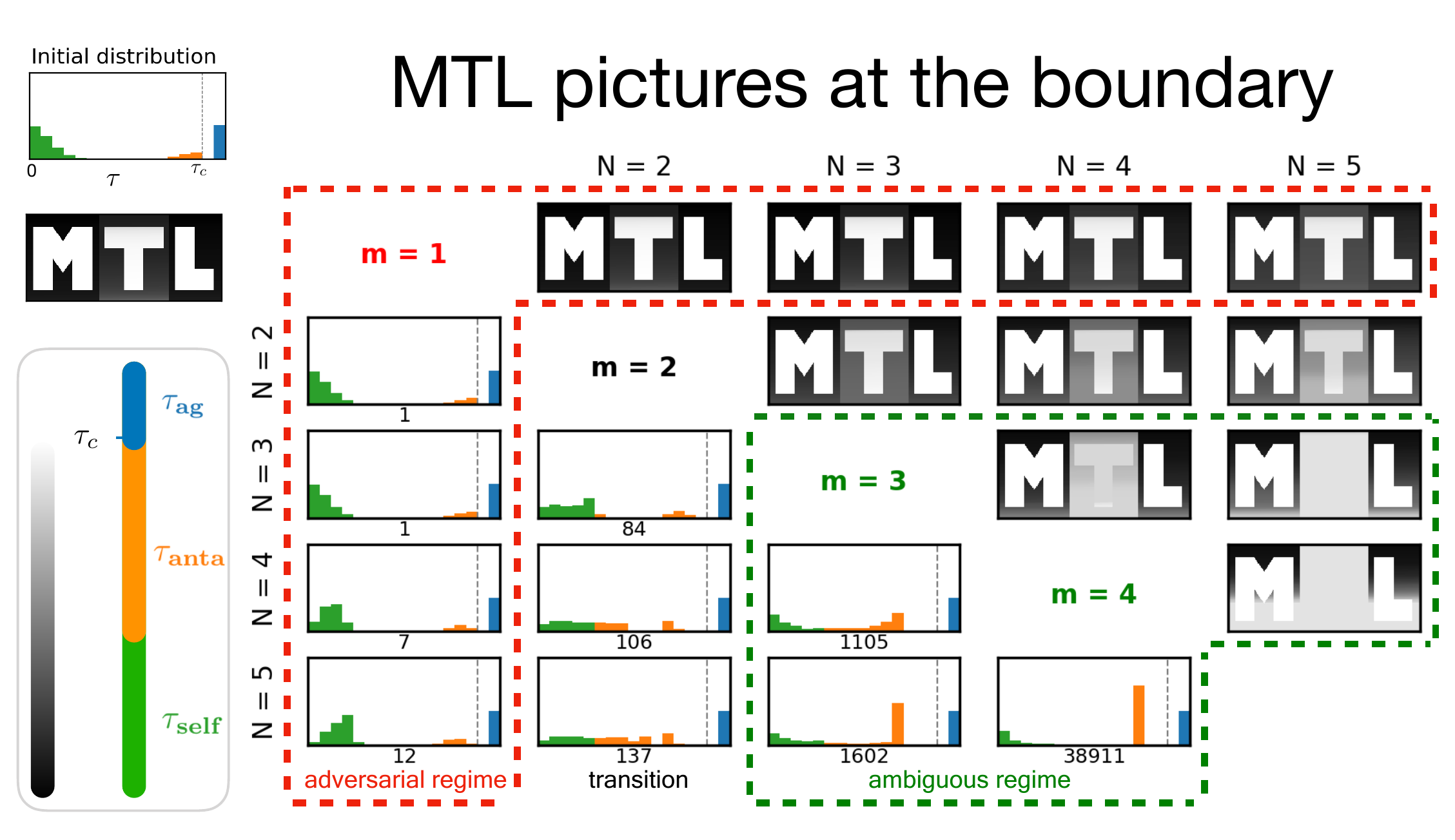}
\caption{\textbf{Figure S\ref*{fig:MTL}.} MTL pictures. Explanation is found in the text}
\label{fig:MTL}
\end{figure*}

We can visually recast immune recognition as an image recognition problem by placing pixels on a grid and coloring them based on their binding time with a given scale. We chose to let white pixels correspond to not self ($\tau > \tau_d$), gray pixels to antagonist ligands ($\tau_a < \tau < \tau_d$) and black pixels to self ligands $\tau \ll \tau_a$. We are free to introduce any kind of spatial correlation to create ``immune pictures'' from a ligand distribution. This results in what we term ``MTL-pictures'' (Fig.~S\ref{fig:MTL}). The initial ligand distribution, MTL picture and scale are given on the left. We perform iterative gradient descent like in the main text, and plot the ligand distribution and the corresponding immune pictures at the boundary for various $(N,m)$. The results are striking. For a T cell operating in the adversarial regime, the ``signal'' MTL is unaltered at the decision boundary. At the transition $m = 2$, we see a slight change of color, while in the ambiguous regime, the signal actually changes from MTL to ML. As we desire for a robust decision-maker, the response should switch when the signal becomes significantly different. From this we conclude, \textit{only in the robust regime can Montreal turn fully into the city of Machine Learning.}

For the MTL pictures in Fig.~S\ref{fig:MTL}, we have distributed the pixels in the $179 \times 431$ frame -- equal to $R$, the number of receptors -- as $L_\text{self} = 0.60R$, $L_\text{a} = 0.12R$, $L_\text{ag} = 0.28R$. We sampled $\tau_\text{self}$ from $|\mathcal{N}(0,\frac{1}{3})|$, $\tau_{\text{a}}$ from  $\tau_d - |\mathcal{N}(0,{1}{3})|$, $\tau_{\text{ag}}$ from $\tau_d + \mathcal{N}(\frac{1}{2},\frac{1}{100})$, and we set $\tau_d = 3$. The picture is engineered such that the agonist ligands fill the M and the L, the antagonists fill the T (which is why the T is slightly darker than the M and L). The self ligands fill the area around the letters M, T and L, such that the self with highest binding time surround the T. We have chosen this example to make the effect of proofreading explicit (and of course because we are based in Montreal and study Machine Learning). This result is generic, and the ambiguity of instances at the decision boundary of a robust model can be visualized with any well-designed image. Scripts to reproduce Fig.~\ref{MTL} A and Fig.~S\ref{fig:MTL} are available at \url{https://github.com/tjrademaker/advxs-antagonism-figs/}.

\subsubsection{Behavior for small binding times}
\label{supp:small-binding-times}
\vspace{-10pt}

Consider a mixture with $L_\text{ag}$ ligands at $\tau_\text{ag} > \tau_d$ and $L$ ligands with small binding time $\tau_\text{spurious} = \tau_\epsilon \ll \tau_\text{ag}$. To understand the behaviour of $T_{N,m}$ as a function of $\tau_\epsilon$ we expand $T_{N,m}$ in small variable $\epsilon = \frac{\tau_\epsilon}{\tau_\text{ag}}$ as
\begin{align*}
T_{N,m}(\{L_\text{ag},\tau_\text{ag};L,\tau_\epsilon\}) &= \frac{ \tau_\text{ag}^N L_\text{ag} + \tau_\epsilon^N L}{ \tau_\text{ag}^m L_\text{ag} + \tau_\epsilon^m L} \\
&= \frac{ 1 + \epsilon^N \frac{L}{L_\text{ag}}}{ 1 + \epsilon^m \frac{L}{L_\text{ag}}} \tau_\text{ag}^{N-m} \\
&\simeq \left(1 + \epsilon^N \frac{L}{L_\text{ag}} \right) \left( 1 - \epsilon^m \frac{L}{L_\text{ag}} \right) \tau_\text{ag}^{N-m} \\
&\simeq \tau_\text{ag}^{N-m} - \tau_\text{ag}^{N-m} \frac{L}{L_\text{ag}} \epsilon^m + O(\epsilon^N),
\end{align*}
which confirms that up to a constant $T_{N,m} \propto -\epsilon^m \propto -\tau_\epsilon^m$ for $m \geq 1$ and $\tau_\epsilon \ll \tau_\text{ag}$, as well as that
\begin{equation}
\frac{dT_{N,m}}{d\tau_\epsilon} \simeq -m \tau_\text{ag}^{N-m-1} \frac{L}{L_\text{ag}} \epsilon^{m-1} \propto -\tau_\epsilon^{m-1}.
\end{equation}

\newpage

%% S4 Appendix
\vspace{-10pt}
\subsection{Appendix S4: Boundary tilting}
\vspace{-10pt}

To further draw the connection between machine learning and adaptive proofreading models, we will study a framework to interpret adversarial examples called boundary tilting \cite{Tanay2016}. We will first illustrate this effect on the discrimination of the original MNIST 3 vs 7 problem MNIST from \cite{Goodfellow2014}) (section \textit{Digit classification}), after which we will interpret boundary tilting via proofreading in ligand discrimination (\textit{Boundary tilting and categorizing perturbations}), and finally, we will derive how the addition of a subthreshold ligand at the decision boundary changes the output (section \textit{Gradient in the $L_2$ direction}). We refer to these results in the main text at the end of section \textbf{Qualitative change in dynamics is due to a critical point for the gradient}.

\subsubsection{Digit classification}
\vspace{-10pt}

\begin{figure*}[!thb]
\centering
\includegraphics[width=\textwidth]{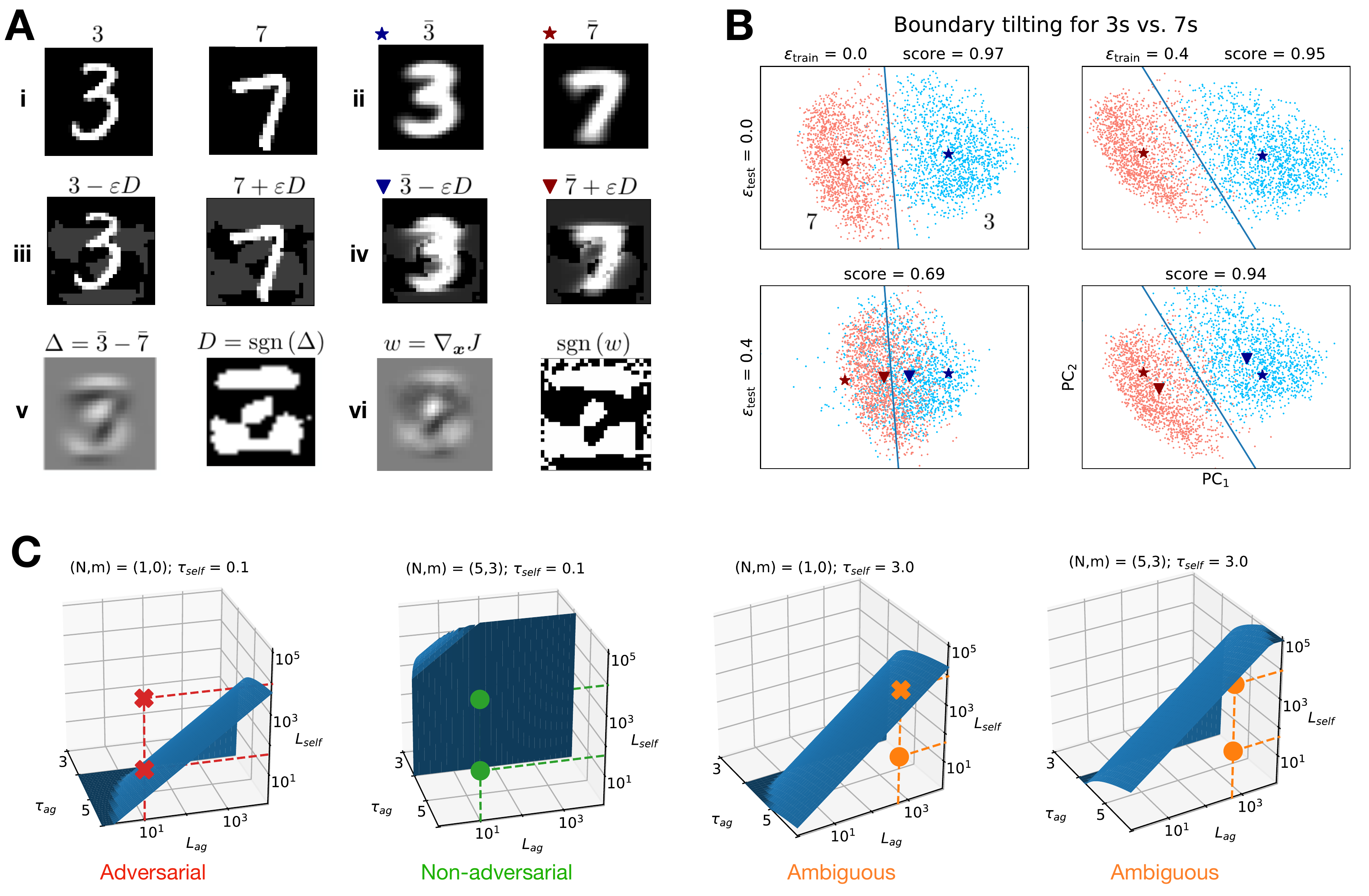}
\caption{\textbf{Figure S\ref*{fig:tilt}.} Boundary tilting in one-dimensional digit classification. (A) (i) Typical 3 and 7 from MNIST. (ii) Average 3, 7 of the traditional test set, (iii, iv) with adversarial perturbation, found by (v) subtracting the sign of $\bar{3}$ from $\bar{7}$, which corresponds to (vi), the perturbation found with FGSM (B) Projection of the digits on the first principal components. The classes are separated by a linear Support Vector Classifier (blue), the average of the classes with and without adversarial perturbation is shown by the triangle and star. We have cycled through permutations of adversarial training and/or adversarial testing. Note how the boundary tilts on the right panels, and how the triangle moves parallel to the decision boundary. (C) Decision boundary of the immune model. The region under the surface is the response regime, the region above is the no-response regime. The classifier with a single proofreading step $(N,m)=(1,0)$ fails to observe agonists in three of the four marked mixtures, while the robust classifier $(N,m)=(5,3)$ correctly responds to each indicated mixture.
}
\label{fig:tilt}
\end{figure*}

A typical 3 and 7 (i), the averages $\bar 3$ and $\bar 7$ (ii), and the corresponding adversarial examples (iii, iv) are shown in Fig.~S\ref{fig:tilt} A. Tanay and Griffin \cite{Tanay2016} pointed out that the adversarial perturbation generated with the Fast Gradient Sign Method (FGSM) proposed in \cite{Goodfellow2014} can also be found via $D = \mathrm{sign} \left( \bar 3 - \bar 7 \right)$, Fig.~S\ref{fig:tilt} A (v). Note the similarity to the adversarial perturbation from the FGSM $\mathrm{sgn}(w) = \mathrm{sgn} \left(\nabla_x J \right)$ (Fig.~S\ref{fig:tilt} A (vi)).
To reveal the linearity of binary digit discrimination, we computed the principal components (PCs) of the traditional training set of 3s and 7s, and projected all digits in the test set on PC$_1$ and PC$_2$ (Fig.~S\ref{fig:tilt} B). With a linear Support Vector Classifier (ordinary linear regression) trained on the transformed coordinates PC$_1$ and PC$_2$ of the training set, we achieve over 95\% accuracy in the test set. While such accuracy is far from the state-of-the-art in digit recognition, it is much higher than typical detection accuracy for single cells (e.g. T cells present false negative rates of 10 \% for strong antagonists \cite{Altan-Bonnet2005}). The red and blue star in Fig.~S\ref{fig:tilt} A denote the average digit $\bar 3, \bar 7$. 

Next, we transformed the test set as $3 \rightarrow 3^\prime = 3 - \epsilon_\text{test} D$, \, $7 \rightarrow 7^\prime = 7 + \epsilon_\text{test} D$, where $\epsilon_\text{test} = 0.4$ is the strength of the adversarial perturbation (Fig.~S\ref{fig:tilt} A (iii)). $\bar{3^\prime}$ and $\bar{7^\prime}$ moved closer in Fig.~S\ref{fig:tilt} B, orthogonal to the decision boundary and along the line between the initial averages. This adversarial perturbation moves the digits in what we call an adversarial direction perpendicular to the decision boundary, and reduces the accuracy of the linear regression model to a mere 69\%. 

Goodfellow et al. proposed adversarial training as a method to mitigate adversarial effects by FGSM. We implemented adversarial training by adding the adversarial perturbation $\epsilon_\text{train} D_\text{train} = \epsilon_\text{train} (\bar 3_\text{train} - \bar 7_\text{train})$ to the images in the training set, computing the new PCs and training the linear regression model. This effectively ``tilts'' the decision boundary, while preserving 95\% accuracy. In the presence of the original adversarial perturbations, we see the effect of the tilted boundary: the perturbation moves digits parallel along the decision boundary, which results in good robust accuracy. This is a illustration example of the more general phenomenon studied in \cite{Tanay2016}.

\subsubsection{Boundary tilting and categorizing perturbations}
\vspace{-10pt}

We consider the change in $T_{N,m}$ for arbitrary $N,m$ upon addition of many spurious ligands. Generalizing Eq. 2 in the main text gives
\begin{equation}
T^\text{after}_{N,m}=\frac{L(\tau-\epsilon)^N+\epsilon R \epsilon^N }{L \tau^m+\epsilon R \epsilon^m}=\frac{(\tau-\epsilon)^N + \frac{ \epsilon^{N+1} R}{L}}{\tau^m+\frac{\epsilon^{m+1} R}{L}}.
\label{eq:anta}
\end{equation}
From this expression, we note that $T_{N,m}$ is changing significantly with respect to its initial value upon addition of many weakly bound ligands as soon as $\epsilon^{m+1} R$ is of order $L$. Thus, the effect described in the main text for weighted averages where $(N,m) = (1,0)$ also holds for nonlinear computations as long as $m$ is small. It appears that the general strategy to defend against this adversarial perturbation is by increasing $m$, as previously observed in \cite{Lalanne2013}. Biochemically, this is done with kinetic proofreading \cite{McKeithan1995,Altan-Bonnet2005,Francois2013}, i.e. we take an output $T_{N,m}$ with $N > m \geq1$.  Here, the output is no longer sensitive to the addition of many weakly bound self ligands, yielding an inversion of the antagonistic hierarchy where the strongest antagonizing ligands exist closer to threshold \cite{Francois2016b}. An extreme case has been proposed for immune recognition where the strongest antagonists are found just below the threshold of activation \cite{Altan-Bonnet2005}. 

We numerically compute how the decision boundary changes when $L_\text{self}$ ligands at $\tau_\text{self}$ are added to the initial $L_\text{ag}$ agonist ligands at $\tau_\text{ag}$, i.e. we compute the manifold so that
\begin{equation}
T_{N,m}(\{L_\text{ag},\tau_\text{ag}; L_\text{self},\tau_\text{self}\})= \frac{ \tau_\text{ag}^N L_\text{ag} + \tau_\text{self}^N L_\text{self}}{ \tau_\text{ag}^m L_\text{ag} + \tau_\text{self}^m L_\text{self}} 
\label{boundary}
\end{equation}
is equal to $T_{N,m}(\{L_\text{ag},\tau_d\})=\tau_d^{N-m}$. We represent this boundary for fixed $\tau_\text{self}$ and variable $L_\text{ag},L_\text{self}, \tau_\text{ag}$ in Fig.~S\ref{fig:tilt} C.  Boundary tilting is studied with respect to the reference $L_\text{self}=0$ plane corresponding to the situation of pure $L_\text{ag}$ ligands at $\tau_\text{ag}$, where the boundary is the line $\tau_\text{ag}=\tau_d$. The case $(N,m)=(1,0)$ (Fig.~S\ref{fig:tilt} C, left panel), corresponds to a very tilted boundary, close to the plane $L_\text{self}=0$, and a strong antagonistic case. In this situation, assuming $\tau_\text{ag} \simeq \tau_d$, each new ligand  added with $\tau_\text{self}$ close to $0$ gives a reduction of $T_{1,0}$ proportional to $\frac{\tau_d}{L_\text{ag}}$ in the limit of small $L_\text{self}$ (see next section, \cite{Francois2016a}), which is again of the order of the response $T_{1,0}=\tau_\text{ag} \simeq \tau_d$ in the plane $L_\text{self}=0$. This is clearly not infinitesimal, corresponding to a steep gradient of $T_{1,0}$ in the $L_\text{self}$ direction. We call the perturbation in this case adversarial. This should be contrasted to the case for higher $m$ (Fig.~S\ref{fig:tilt} C, middle left) where the boundary is vertical, independent of $L_\text{self}$, such that decision-making is based only on the initially present $L_\text{ag}$ ligands at $\tau_\text{ag}$. Here, the change of response induced by the addition of each ligand with small binding time $\tau_\text{self}$ is $\tau_\text{self}^m$, due to proofreading a very small number when $\tau_\text{self} \simeq 0$ \cite{Francois2016a}. Contrary to the previous case, the gradient of $T_{N,m}$ with respect to this vertical direction is almost flat and very small compared to the response in the $L_\text{self}=0$ plane. We call the perturbation in this case non-adversarial. 

Tilting of the boundary only occurs when $\tau_\text{self}$ gets sufficiently close to the threshold binding time $\tau_d$ (Fig.~S\ref{fig:tilt} C, right panels). In this regime, each new ligand added with quality $\tau_\text{self}=\tau_d-\epsilon$ contributes an infinitesimal change of $T_{N,m}$  proportional to $\frac{\tau_d-\tau_\text{self}}{L_\text{ag}}=\epsilon/L_\text{ag}$, which gives a weak gradient in the direction $L_\text{self}$. But even with such small perturbations one can easily cross the boundary because of the proximity of $\tau_\text{self}$ to $\tau_d$, which explains the tilting. The cases where the boundary is tilted and the gradient is weak are of a different nature compared to the adversarial case of Fig.~S\ref{fig:tilt} C, left panel. Here the boundary is tilted as well, but the gradient is steep, not weak. For this reason we term the cases on the right panels ambiguous. Similar ambiguity is observed experimentally: it is well known that antagonists (ligands close to thresholds) also weakly agonize an immune response \cite{Altan-Bonnet2005}. Our categorization of perturbations is presented in Table~S\ref{tab:definitions}. Scripts for boundary tilting in ligand discrimination and digit discrimination are available at \url{https://github.com/tjrademaker/advxs-antagonism-figs/}.

%In conclusion, with adversarial training, we have tilted the decision boundary and removed an adversarial direction, providing a simple and striking example of \cite{Tanay2016}, and illustrates beautifully how proofreading removes the adversarial direction along the $L_2$ axis in Fig. \ref{fig:FGSM} C of the main text. 

\begin{table}
\centering
\caption{\textbf{Table S1}: Categories of perturbations}
\label{tab:definitions}
\begin{tabular}{ccc}
\toprule
 & Boundary tilting & Gradient when adding\\& & one antagonistic ligand \\
Adversarial & yes & steep ($\mathcal{O}(1)$) \\
Non-adversarial & no & almost flat ($\mathcal{O}(\epsilon^m)$) \\
Ambiguous & yes & weak ($\mathcal{O}(\epsilon)$)
\end{tabular}
\medskip
\end{table}

\subsubsection{Gradient in the $L_2$ direction}
\label{supp:grad}
\vspace{-10pt}

We recall results from \cite{Francois2016b} to show how the addition of subthreshold ligands one at a time changes the output. We first consider $\{ L, \tau_d \}$ threshold ligands with output
\begin{equation}
T_{N,m}(L,\tau_d) = \tau_d^{N-m}.
\end{equation}
The main result of \cite{Francois2016b} is the linear response of $T_{N,m}(L,\tau_d)$ to the addition of $\{ L_a, \tau_d - \epsilon \}$ subthreshold ligands.
\begin{align}
&T_{N,m} \left( \{ L, \tau_d; L_a, \tau_d - \epsilon \} \right) \\
&= T \left( L + L_a, \tau_d \right) - \epsilon L_a \mathcal{A} \left( L + L_a, \tau_d \right) \\
&= \tau_d^{N-m} - \epsilon \frac{L_a}{L+L_a} \frac{ d }{ d\tau } T_{N,m}(L+L_a,\tau) \Big|_{\tau = \tau_d},
\label{eq:mix}
\end{align}
where we used the definition
\begin{equation}
\mathcal{A} \left( L, \tau_d \right) = \frac{1}{L} \frac{ d }{ d\tau } T_{N,m}(L,\tau) \Big|_{\tau = \tau_d}.
\end{equation}
for the coefficient in a mean-field description. As the derivative $\frac{ d }{ d\tau } T_{N,m}(L,\tau)  \Big|_{\tau = \tau_d} > 0$, and $\epsilon = \tau_a - \tau_d$, each additional subthreshold ligand at $\tau_a$ decreases the output with a value proportional to
\begin{equation}
\frac{ \tau_d - \tau_a }{ L }.
\end{equation}
In the case $(N,m) = (1,0)$, the mean-field approximation is exact, i.e. the first derivative of $\frac{dT}{d\tau}$ is the only nonzero derivative, given by
\begin{equation}
\mathcal{A}(L,\tau_d) = \frac{1}{L} \frac{d}{d\tau} \tau \Big|_{\tau=\tau_d} = \frac{1}{L}.
\end{equation}
With the addition of a single subthreshold ligand $\tau_a \simeq 0$, so that $\epsilon \simeq \tau_d$, the output is maximally reduced by $\frac{\tau_d}{L + 1} \simeq \frac{\tau_d}{L}$, a finite quantity, as described in the main text. For higher m, the linear approximation holds only for ligands at $\tau_a$ close to threshold. 

% As $m$ increases, we have $\epsilon = \tau_d - \tau_2$ becoming infinitesimally small with $\tau_2 \to \tau_d$, corresponding to a smooth gradient in the adversarial direction $L_2$. 

\newpage

%% S5 APPENDIX SECTION
\vspace{-10pt}
\subsection{Appendix S5: Few-pixel attack}
\label{supp:few-pixel-attack}
\vspace{-10pt}

In Appendix S5, we describe in detail the procedure for the few-pixel attack. We used this to come to our conclusion in section \textbf{Biomimetic defenses against few-pixel attacks} and Fig.~\ref{MTL} C in the main text.

The few-pixel attack connects to ligand antagonism in the sense that few pixels are needed to cause misclassification, corresponding to the addition of few maximally antagonizing ligands to a mixture fooling robust adaptive proofreading models. It is not the most efficient attack against a classifier without biomimetic defence, but it is the most efficient attack against classifiers with biomimetic defence, equivalent to adaptive proofreading models with $m>1$. For these adaptive proofreading models, there exists a unique maximally antagonistic binding time, defined as the binding time that maximally reduces $T_{N,m}$.

\begin{figure*}[!thb]
\centering
\includegraphics[width=0.75\textwidth]{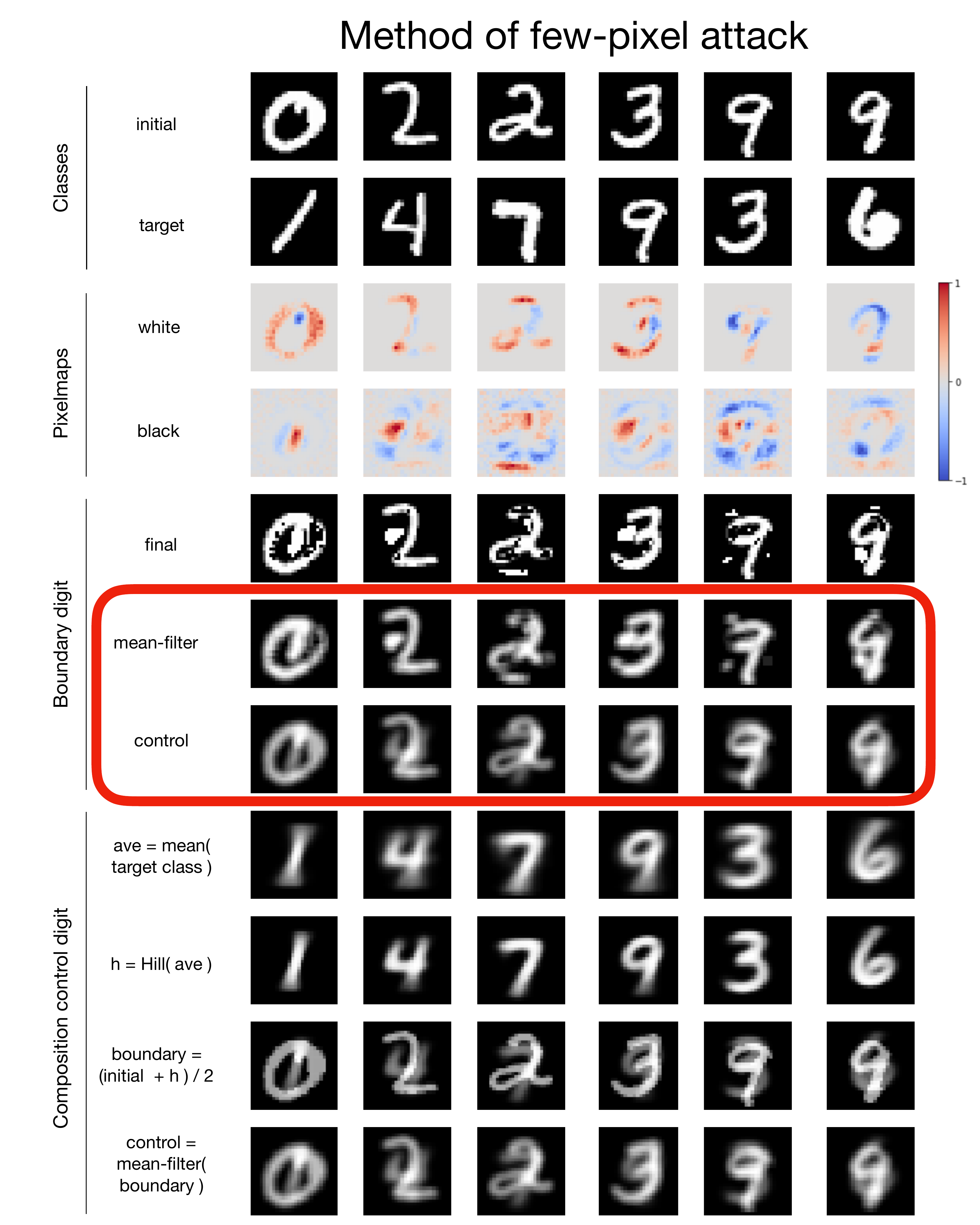}
\caption{\textbf{Figure S\ref*{fig:few-pixel-attack}.} Method of few-pixel attack. Each column show how a few-pixel attack causes misclassification of an initial digit to a target class. The important result are the pre-filtered boundary digits and the control in the red rectangle. Pixelmaps determine which pixels increase (red) or decrease (blue) the score when turning an individual pixel in the initial digit white or black. We merge the pixelmaps, sort this list of pixels, and go through it from maximum to minimum change in score until misclassification occurs, resulting in the pre-filtered digit. We apply a mean-filter to make them look more like real digits, and indeed, these mean-filtered boundary digits closely resemble our control digits at the boundary. The control digits are composed of the mean-filtered initial digit plus locally contrasted (with hill function ($N = 3; \theta = 0.5$) average digit of the target class.}
\label{fig:few-pixel-attack}
\end{figure*}

With this in mind, we decided to make pixels black or white in a controlled manner, until the neural network classifies the perturbed, initial digit as the target class. In the following, we will refer to several stages of the few-pixel attack using Fig.~S\ref{fig:few-pixel-attack}. We first computed what we term pixelmaps. Pixelmaps contain the change of score when making a pixel white or black. In Fig.~S\ref{fig:few-pixel-attack}, blue colors correspond to pixels that will lower the score when turned white or black, while red colors are for pixels that will increase the score for the same operation. A grey color means the score is unchanged when whitening or blacking the pixel. The pixelmaps are scaled to the maximum change in score. We proceed in merging and sorting the pixelmaps from maximum to minimum change in score towards the target class, iteratively following the sorted list to decide which pixels in our digit to turn white or black. We do this until we reach the decision boundary (first iteration in which the digit is misclassified). The final digits in the row above the red rectangle in Fig.~S\ref{fig:few-pixel-attack} are the resulting boundary digits. They already contain perturbations corresponding to real features, but have an air of artificiality to them which allows us to fairly easily distill the ground truth. We remove this with a mean filtering \cite{Xie2018}, which is a 3x3 convolutional block that computes mean pixel values as
\begin{equation}
y_{i,j} = \frac{1}{9} \sum_{k,l = -1} ^1 x_{i+k,j+l}.
\end{equation}
Biologically, this is pure receptor clustering, where a perturbation to a single receptor locally affects other ligands. Such digits are truly ambiguous digits that are tough to classify even as humans. These are the type of digits we expect to find on the decision boundary. Finally, we compare the mean-filtered digit at the decision boundary to the control: the sum of the initial digit and the hill function of Eq. \ref{biomimetic} ($N = 3; \theta = 0.5)$ on the average of all digits in the target class, then mean-filtered (Fig.~S\ref{fig:few-pixel-attack} for a step-by-step composition). We apply the mean-filter to the control to again remove the artificiality of a digit plus an average, and make the comparison between boundary digit and control digit fairer. The similarity between mean-filtered boundary digit and control digit confirms our intuition that we are actually operating in the space between both classes when misclassification occurs.

\begin{figure*}[!thb]
\centering
\includegraphics[width=0.75\textwidth]{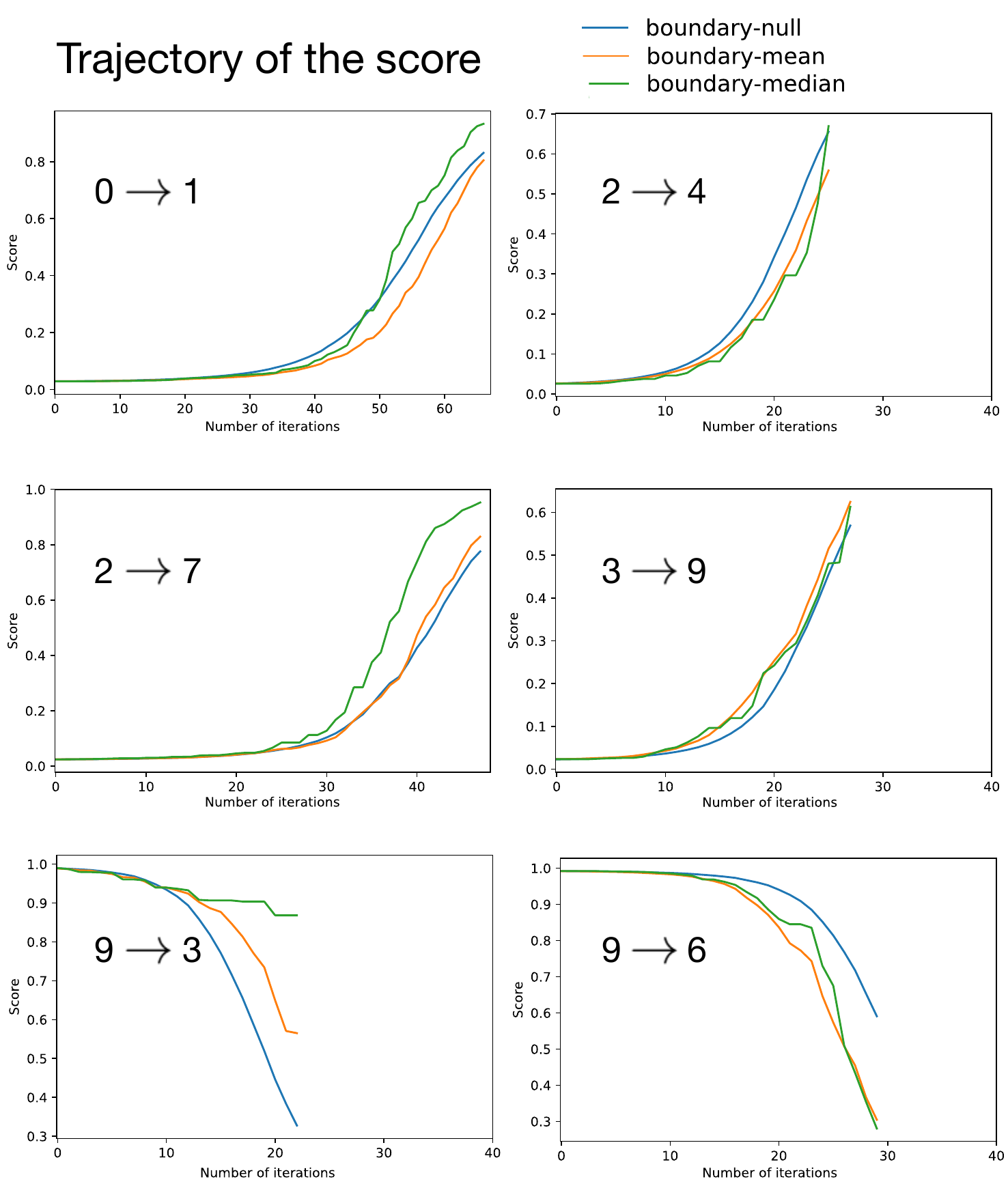}
\caption{\textbf{Figure S\ref*{fig:score-few-pixel-attack}.} Trajectory of the scoring functions of the attacks in Fig.~S\ref{fig:few-pixel-attack}. The blue, orange and green line correspond to various digits (actual digit, mean-filtered digit, median-filtered digit) for which we check the score, and terminate when reaching the boundary. The trajectory of the score for the null digit and the mean-filtered digit is generally the same. Moreover, the behavior of the score looks similar to the behavior of $T_{N,m}$ upon addition of maximally antagonizing ligands to a mixture of only agonist ligands in Fig.~\ref{fig:FGSM} D in the main text.}
\label{fig:score-few-pixel-attack}
\end{figure*}

We can also apply the mean-filter to the initial digit before generating the pixelmaps, and during the procedure, check the score on the mean-filtered perturbed image. This gives similar results, as we see by following the trajectory of the score for \textit{boundary-null} and \textit{boundary-mean}. We have shown the score explicitly in Fig.~S\ref{fig:score-few-pixel-attack} for the digits in Fig.~S\ref{fig:few-pixel-attack}. The behavior of the score is remarkably similar to the interpolation between ligand mixtures (Fig.~\ref{fig:FGSM}F, bottom panel in the main text). A nonlinear filtering method proposed in \cite{Xie2018} is the median-filter, but this one works less well for black-and-white pixels.

We have shown examples that are generated when we select for instances where the number of iterations is large enough (20 suffices, we still consider this to be a few-pixel attack, keeping in mind that digits have 784 individual pixels). The authors of \cite{Su2017} specifically searched for single pixel attacks. Examples of single-pixel misclassification exist in our neural networks trained on two types of digits in MNIST too, but these we found non-informative. In cellular decision-making, this case corresponds to adding a single antagonist ligand to a ligand mixture to cause misclassification. This is only possible if the ligand mixture is already very close to the boundary. For such samples, we do not expect ambiguity to appear. Remember that near the boundary, the score landscape is steep, and small additions have a large effect.

\end{document}